\def\year{2021}\relax
\documentclass[letterpaper]{article} % DO NOT CHANGE THIS
\usepackage{aaai21}  % DO NOT CHANGE THIS
\usepackage{times}  % DO NOT CHANGE THIS
\usepackage{helvet} % DO NOT CHANGE THIS
\usepackage{courier}  % DO NOT CHANGE THIS
\usepackage[hyphens]{url}  % DO NOT CHANGE THIS
\usepackage{graphicx} % DO NOT CHANGE THIS
\usepackage{natbib}  % DO NOT CHANGE THIS AND DO NOT ADD ANY OPTIONS TO IT
\usepackage{caption} % DO NOT CHANGE THIS AND DO NOT ADD ANY OPTIONS TO IT
\title{Understanding the Dynamics between Vaping and Cannabis Legalization\\ Using Twitter Opinions}
\author {
    Shishir Adhikari,\textsuperscript{\rm 1}
    Akshay Uppal,\textsuperscript{\rm 1}
    Robin Mermelstein,\textsuperscript{\rm 2}
    Tanya Berger-Wolf,\textsuperscript{\rm 1,3}%\thanks{A part of the work was performed while at UIC.}
    Elena Zheleva\textsuperscript{\rm 1} \\
}
\affiliations {
    \textsuperscript{\rm 1}Computer Science, University of Illinois at Chicago \\
    \textsuperscript{\rm 2}Psychology; Institute for Health Research and Policy, University of Illinois at Chicago \\
    \textsuperscript{\rm 3}Computer Science and Engineering; Electrical and Computer Engineering; Evolution, Ecology, and Organismal Biology; Translational Data Analytics Institute, Ohio State University \\
    sadhik9@uic.edu, auppal8@uic.edu, robinm@uic.edu, berger-wolf.1@osu.edu, ezheleva@uic.edu \\
}

\usepackage{amsmath}
\usepackage{array}
\usepackage{multirow}
\usepackage{subcaption}
\usepackage[inline]{enumitem}
\newcommand{\norm}[1]{\left\lVert#1\right\rVert}

\begin{document}

\maketitle

\begin{abstract}
Cannabis legalization has been welcomed by many U.S. states but its role in escalation from 
tobacco e-cigarette use to cannabis vaping is unclear. 
Meanwhile, cannabis vaping has been associated with new lung diseases and rising adolescent use.
To understand the impact of cannabis legalization on escalation, we design an observational study to estimate the causal effect of recreational cannabis legalization on the development of pro-cannabis attitude for e-cigarette users. 
We collect and analyze Twitter data which contains opinions about cannabis and JUUL, a very popular e-cigarette brand. We use weakly supervised learning for personal tweet filtering and classification for stance detection.
We discover that recreational cannabis legalization policy has an effect on increased development of pro-cannabis attitudes for users already in favor of e-cigarettes.
\end{abstract}

\section{Introduction}
Electronic cigarettes (e-cigarettes) are battery-powered devices designed to heat up a liquid containing nicotine, cannabis and other chemicals, and to deliver its contents in the form of an aerosol~\cite{cullen2018notes}. 
Vaping is the process of inhaling aerosols from an e-cigarette. Studies show that e-cigarettes have the potential to benefit adult smokers as a tobacco smoking cessation tool~\cite{hajek-nejm19,walker-lrm20}. Although e-cigarettes are considered to be less harmful than smoking combustible tobacco products like cigarettes and cigars, e-cigarettes have risks associated with respiratory health~\cite{polosa-errm19} and nicotine addiction~\cite{barrington-jah16}. 
E-cigarettes are also widely blamed for the recent increase of tobacco use in adolescents due to targeted advertising, appealing flavors, social normalization, lack of awareness about health risks, illegal sales, and easy access to minors~\cite{jackler-srita19,barrington-pediatrics15,fda-online18}. 
This increase has been largely attributed to JUUL, an e-cigarette with a discreet USB-like design, which dominates the U.S. e-cigarette market with $82.9\%$ market share as of 2019~\cite{conway-statista20}. 
JUUL and other e-cigarette companies have been the target of recent policy regulations with the goal of mitigating adolescent tobacco use~\cite{jackler-srita19}. The U.S. Surgeon General report concluded that the growing popularity of e-cigarettes among minors
is a public health concern and suggested plans for spreading health risk awareness and market regulations~\cite{usdhhs-nccdphp16}. 

Concomitant with the growing use of e-cigarettes has been the increasing use of cannabis vaping~\cite{mtf-2020}. 
Cannabis vaporization among adult cannabis users is believed to be a healthier alternative than combustible cannabis, with better taste and weaker odor, flexibility in carrying and concealing, and better euphoric sensation~\cite{morean-sum17}. The psychoactive property of cannabis attributed to its Tetrahydrocannabinol (THC) content has attracted medical and recreational use resulting in some US states legalizing cannabis use.
Despite its positive traits, cannabis vaping has also raised concerns. From 2017 to 2019, cannabis vaping has doubled to tripled amongst high school students~\cite{mtf-2020} and the use of THC products is linked to the 2019 nationwide outbreak of e-cigarette or vaping product use–associated lung injury (EVALI)~\cite{chatham-mmwr19}. The cannabinoid-enriched e-liquids readily available in the market lack quality control as well as toxicological and clinical assessment~\cite{giroud-ijerph15}. This controversy poses policy-related questions for states looking to legalize cannabis. 

Although researchers are still evaluating the effects of cannabis legalization legislation on patterns of cannabis use and associated harms across different populations~\cite{smart-ajdaa19}, current tobacco and e-cigarette users may be an especially vulnerable group for cannabis use escalation and related harms~\cite{agrawal-a12,lemyre-sum19}, including continued tobacco use and increased dependence and difficulty in stopping use. Thus policy makers and public health professionals may need to think about targeted messages and interventions to help discourage co-use of substances and to increase tobacco cessation interventions for these more vulnerable groups. 

Recent studies have also found that cannabis legalization laws may have an association with modes of cannabis administration. In both adolescents and adults, individuals who live in states with legal cannabis laws are more likely to try vaping cannabis than those who do not reside in states with legalized cannabis~\cite{borodovsky-ijdp16,borodovsky-dad17}. We do not yet know, though whether there is a causal or even temporal association between legalization and mode of use, but examining whether legalization is associated with how cannabis is consumed can have important policy and health implication.  

The goal of our work is to understand the potential association between cannabis legalization and increases in cannabis vaping interest (and potential use) among individuals already predisposed to vaping e-cigarettes or electronic nicotine delivery systems.
Even though studies~\cite{dai-pediatrics18,audrain-pediatrics18,giroud-ijerph15} have looked into e-cigarette and cannabis co-use, the relationship between the policy of legalizing recreational cannabis and this co-use is not well-understood.  
Social media platforms have been used to share personal experiences and opinions related to cannabis and e-cigarettes use~\cite{cavazos-jmir14,kim-jmir17}. However, most of the studies rely on survey data collected from volunteers and do not utilize the vast amount of social media content.  

To estimate the causal effect of recreational cannabis legalization on the development of pro-cannabis attitudes and potential cannabis vaping in e-cigarette users, we design an observational study using Twitter data and matching~\cite{stuart-sc10}. Our population of interest is U.S. users who have expressed positive stance towards JUUL vaping but have not mentioned cannabis before a focus state's recreational cannabis legalization date.
We assign the users from that state (e.g., California) to the treatment group and the other users are assigned to four control groups based on their state's cannabis policies, ordered from most to least strict:
(1) Users from states with illegal cannabis, (2) Users from states with legal medical cannabis with limited THC, (3) Users from states with legal medical cannabis, and (4) Users from states with legal recreational cannabis. The outcome is whether a user initiates a pro-cannabis tweet within $N$ months of the legalization effective date of the treatment state. Since the treatment and control assignment is not randomized, we use the potential outcome framework to create a balanced pseudo-population for estimating the causal effect of the treatment on the outcome.

In order to deal with the noisy Twitter data, our methodology involves multiple steps. We collect from Twitter a dataset of JUUL users and their vaping- and cannabis-related tweets. We utilize weakly supervised learning to identify tweets related to personal experiences, opinions, and observations. Focusing on individuals with personal tweets allows us to study the topics of interest for ordinary users. We annotate a subset of the tweets asking annotators to label the stance of the tweet author on e-cigarettes or cannabis use to one of three classes: ``In favor," ``Against," or ``Neither." We use supervised stance detection to predict stance in the full dataset. 
In contrast to the sentiment analysis methods that focus on determining the polarity of emotion expressed in a text, stance detection is concerned with identifying the position or standing of users on a particular topic. Finally, to account for confounding, we use matching to find the appropriate control group users and estimate the effect of cannabis legalization on cannabis vaping rates of e-cigarette users.

\section{Background and Related Work}
\textbf{Social media data in public health.} Previous studies have used data from social media for the analysis of public health issues such as maternal mortality, pregnancy, mental health, and substance use~\cite{abebe-www20,liu-www19,coppersmith-acl14,cavazos-jmir14,kiciman-icwsm18}. Social media data has also been used to study cannabis user behaviors and online interactions~\cite{cabrera-nguyen-jsad16,cavazos-jmir14}. 
\citet{kim-jmir17} used e-cigarette-related tweets and user profile information to categorize users into individual, vape enthusiasts, informed agencies (news or health), marketer, and spammer. \citet{unger-pm18} performed a study to show an association between tobacco-related Twitter activities and actual tobacco use. There have been studies on understanding the public response to public health policy changes using content analysis of a few thousand tweets~\cite{hatchard-plos19,harris-jmir14}. Although previous work~\cite{young-fph19} outlines the potential of social media for understanding the impact of the cannabis legalization policy, there has been no study on estimating the causal effect of recreational cannabis legalization policy on the co-use of cannabis with e-cigarettes and our study is the first to do that.

\textbf{Opinion mining and stance detection.}
Opinion mining is often used interchangeably with the sentiment analysis task that mainly focuses on opinions expressing or implying positive or negative sentiments~\cite{liu-slhlt12}. The machine learning approach to sentiment analysis requires labeled data to train sentiment classifiers whereas the lexicon-based approach uses predefined rules and generic polarity scores of words expressing sentiment. VADER~\cite{hutto-aaai14} is a popular lexicon-based sentiment analysis tool designed for sentiment analysis of unlabeled social media text by handling slangs and emoticons. However, VADER focuses on the tweet polarity and it is insensitive to the target context (attitude towards cannabis or vaping). 
\citet{aldayel-icsi19} distinguish sentiment analysis from stance detection, a task of inferring a supportive or opposing attitude towards a given topic or entity. \citet{aldayel-arxiv20} survey the literature of stance detection on social media and identify three main machine learning approaches: supervised learning, weakly-supervised with transfer learning, and unsupervised learning.

\textbf{Causal effect estimation.} 
\citet{rubin-jep74} formalized the potential outcome framework for causal effect estimation from randomized and non-randomized studies. In an observational study with non-randomized treatment assignment, matching methods can be employed for balancing the covariates between treatment and control groups~\cite{stuart-sc10}. Sensitivity analysis on the matching parameter choices is recommended to provide less subjective results~\cite{king-pa19,shahid-why19}.
For the estimation of the variance of the causal effect, bootstrapping is recommended~\cite{stuart-sc10,austin-sm16}. Matching methods have been used in answering public health and socio-economic questions using social media data~\cite{dos-aaai15,de-icwsm17,altenburger-icwsm17,park-icwsm20}. Quasi-experimental designs have also been used to handle unobserved confounding in policy-related observational studies with social media data~\cite{tian-aaai20}.

\section{Problem Description}

The main goal of our study is:
\begin{quote}
     \emph{To \textbf{understand the impact} of recreational cannabis legalization on the development of \textbf{pro-cannabis attitude} for the population with a \textbf{positive attitude} toward e-cigarettes by analyzing \textbf{public opinions} on Twitter.}
\end{quote}
 
To achieve this goal, we collect and thoroughly analyze a new dataset with JUUL and cannabis-related tweets. 
Our main task is \emph{to estimate the causal effect} of legalization on developing a pro cannabis attitude for e-cigarette users. We define three additional tasks to support the data preprocessing and analysis.

% Table: Dataset
\begin{table*}[!t]
    \centering
    \begin{tabular}{|c|c|c|c|m{9cm}|}
    \hline
        \textbf{Dataset} & \textbf{Date Filter} & \textbf{\# Users} & \textbf{\# Tweets} & \textbf{Hashtags or Keywords Filter} \\
    \hline
         $D_J$ & 2016-2018 & 312K & 597K & `juul', `juulvapor', `juulnation', `doit4juul' \\
 \hline
         $D_C$ & 2014-2018 & 194K & 3.28M & `weed', `ganja', `marijuana',  `cannabis', `mary jane', `THC', `marihuana', `hash', `reefer', `hashish', `bhang', `CBD', `green goddess', `locoweed', `maryjane', `spliff', `hemp', `wacky baccy', `sinsemilla', `doobie', `acapulco gold'\\
         \hline
    \end{tabular}
    \caption{Summary of the dataset collected from Twitter. The number of users and tweets are for the dataset after applying language and location filters.
    }
    \label{tab:collection}
\end{table*}

\subsection{Data Collection and Cleaning} 
\label{sec:dataset}

Twitter is a social media platform skewed towards younger users \cite{TwitterAge} which makes it appropriate for our study.
Since JUUL is the dominant e-cigarette of choice, we focused on collecting tweets related to JUUL from its inception in 2016 to 2018 using Crimson Hexagon API\footnote{https://apidocs.crimsonhexagon.com/}. We also collect the cannabis-related tweets for those users from 2014 to 2018. We use keyword and hashtag filters provided by a domain expert on the team, summarized in Table \ref{tab:collection} along with statistics of the number of users and tweets selected in the final JUUL-related dataset $D_J$ and cannabis-related dataset $D_C$. The annotations described in more detail in the section on Stance Detection showed that this keyword strategy has a very low false positive rate of $0.5\%$ and $3.8\%$ in the annotated samples from $D_J$ and $D_C$ respectively.

\begin{figure*}[!ht]
    \centering
    \includegraphics[width=\textwidth, keepaspectratio]{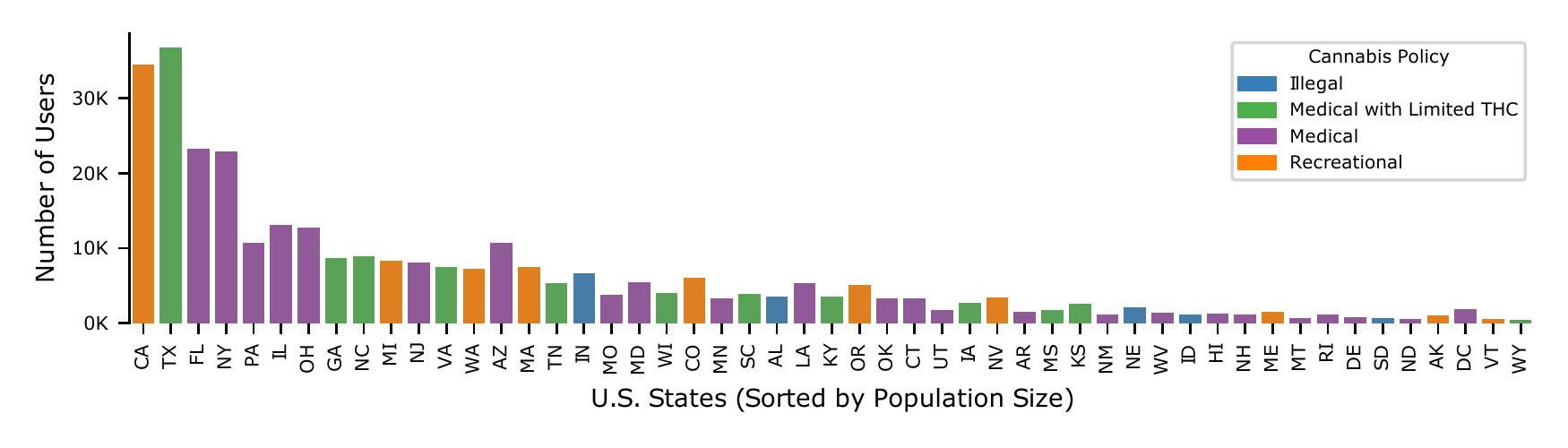}
    \caption{Distribution of users in dataset $D_J$ with JUUL-related tweets across the U.S. states along with the cannabis policies of the states as of December 2018}
    \label{fig:state_dist}
\end{figure*}

From the collected raw tweets, we retain only English tweets of U.S. users, discarding all tweets from JUUL's official Twitter handle. Using regular expressions, we match the user profile's reported location to the fifty U.S. states and the District of Columbia. Since the location is optionally entered by the users as free text, we discard users with missing location information or users with other text string unrelated to location. There is also a variance in the location format for the users that have entered the location. We match the self-reported location to U.S. address patterns containing at least state name, state abbreviation, or main cities of the states. 
Figure \ref{fig:state_dist} shows the distribution of users with JUUL-related tweets across the U.S. states along with the cannabis policies of the states as of December 2018.

\subsection{Data Model}
We represent the data as two types of entities, tweets and users.

\textbf{Tweet entity.}
A tweet instance $M_i \in \mathbf{M}$ is defined in terms of its attributes:
$d$ is the tweet creation date, $W$ is the raw tweet text, $\mathbf{X}$ is a vector representation of tweet text, and $v$ is the ID of user posting the tweet. $\mathbf{X}$ is represented with the mean word vector obtained from pre-trained Glove Twitter word embeddings~\cite{pennington-emnlp14}.

\textbf{User entity.}
A user instance $V_i \in \mathbf{V}$ is defined in terms of its attributes which can be divided into two categories: raw and aggregated. Raw attribute consist of user's U.S. address ($s$). We also consider other raw user attributes such as friend counts, follower counts, number of statuses liked but discard those since the values of these covariates obtained from Twitter API are not guaranteed to have been measured pre-treatment as required by the causal estimation task.
The first aggregated attribute is the mean of tweet text representation posted by the user {\small $v_i.X = mean(m_i.X \mid m_i.v = v_i)$}. The other aggregated attributes are the first JUUL-related tweet creation date {\small $v_i.d_j = min(m_i.d \mid m_i.v = v_i \wedge \mathbf{M} \in D_J)$}, and the first cannabis-related tweet creation date {\small $v_i.d_c = min(m_i.d \mid m_i.v = v_i \wedge \mathbf{M} \in D_C)$}.

\subsection{Data Analysis Tasks}

\textbf{Personal tweet selection.} Since we are interested in the opinions of the general public, the first task is to identify personal tweets. We refer to a tweet as a "personal" tweet if the tweet author shares experiences, opinions, or observations on cannabis/e-cigarette use, whether their own or someone else's. All other tweets are referred to as ``non-personal" tweets. Non-personal tweets can consist of promotional content, factual information or news. We frame the personal tweet selection problem as a binary classification task and retain the tweets belonging to the personal class for further analysis. We consider both tweets and retweets because previous work has shown that retweeting usually indicates trust and agreement with the tweet content~\cite{metaxas-icwsm15}. We also present results without the retweets.

\textbf{Topic analysis.} The second task is to gain understanding of the different JUUL-related and cannabis-related topics shared by users in their personal tweets. Topic analysis gives intuition about the breadth of expressed opinions and experiences and their relevance to the research goal. The topic analysis also helps us perform a qualitative evaluation of the personal tweet selection process.

\textbf{Stance detection.} The third task is to identify whether a user is in favor of or against cannabis and e-cigarette use by detecting stance in personal tweets. A user can express a positive attitude toward e-cigarette or cannabis by tweeting about ongoing personal use, the experience of use, intention to use, positive opinion or advantages of e-cigarettes or cannabis. Similarly, a negative outlook on e-cigarette or cannabis can be expressed by tweeting negative experiences, negative opinions, intention to quit, information about harmful effects or disadvantages. We perform stance detection for each tweet by determining whether the tweet author is in favor of or against e-cigarette or cannabis use. Tweets can belong to one of three classes: "In favor," "Against," or "Neither." We define a user as pro-e-cigarettes or pro-cannabis if the majority of the tweets posted by the user in a given period are in support of JUUL or cannabis use.

\subsection{Causal Effect Estimation}
Our main task is to estimate the causal effect of recreational cannabis legalization policy on the initiation of pro-cannabis tweets for the pro-e-cigarettes population. The potential outcome framework considers each unit to have two potential outcomes depending on whether the unit is treated. The causal effect of a treatment ({\small $T \in \{0, 1\}$}) is defined as the difference between the two potential outcomes,
{\small $v_i.E = v_i.Y(1) - v_i.Y(0)$}, where {\small $v_i.Y(1)$} is the user $v_i$'s outcome when treated, {\small $v_i.Y(0)$} the outcome when not treated. Due to the fundamental problem of causal inference, the causal effect is generally calculated at the population level. The average treatment effect (ATE) is obtained as the difference of mean outcome between treatment and control groups. The gold standard for causal effect estimation is randomized controlled trials (RCT) which randomly assign the population of interest into treatment and control groups. Randomization ensures that confounding cannot impact the estimate and that unit covariates are balanced. When RCTs are not possible, researchers resort to quasi-experimental design~\cite{shadish-book02}. In our work we rely on a specific type of such design, called matching. Matching creates a pseudo-population to balance the covariate distribution in the treatment and control group~\cite{stuart-sc10}.
In Nearest Neighbor Matching (NNM), each instance in the treatment group $(v_i)$ is matched with the nearest neighbor in the control group $(v_j)$ based on a distance metric $(D)$ that takes covariates as input. Mathematically, the matched pair $(v_i, v_j)$ is defined as follows:
{ \small
\begin{equation}
\begin{split}
  (v_i, v_j) | D(v_i.X, v_j.X) \leq D(v_i.X, v_k.X)~\wedge~v_i.T=1~\wedge\\v_j.T=0~\wedge~v_k.T=0~\wedge~v_i,v_j,v_k \in \mathbf{V} \wedge v_i \neq v_j \neq v_k
  \end{split}
\end{equation}
}where $v_i.T=1$ indicates the user $v_i$ is in treatment group.
In Propensity Score Matching (PSM), the covariates are summarized by a scalar that acts as a distance measure to choose appropriate control subjects. Propensity scores can also used in Inverse Probability of Treatment Weighting (IPTW) which inverse-weighs treatment and control subjects for creating a balanced pseudo-population. A user's propensity score ($v_i.e$) is defined as its probability of receiving the treatment given its covariates i.e. {\small $v_i.e = P(v_i.T = 1 | v_i.X)$}.
The estimation of ATE with matching works with three main assumptions: ignorability, stable unit treatment value assumption (SUTVA), and positivity. Ignorability states that the treatment assignment of a unit is independent of the potential outcomes conditional on the observed covariates of the unit. This assumption implies that there are no unobserved confounders. The SUTVA assumption states that the outcomes of a unit are not affected by the treatment assignments of other units.
Positivity states that the probability of receiving either treatment by each unit is larger than zero.

\section{Data Analysis}

In this section, we describe the methodologies employed for selecting personal tweets, topic extraction, stance detection, and the results they yield on our dataset.

\subsection{Personal Tweet Selection}
We pose the task of identifying personal tweets as a binary classification problem with two classes: ``personal" and ``non-personal." We apply a weakly supervised approach to classify personal tweets and then filter out tweets from accounts that have predominantly non-personal tweets together with tweets from potential bot accounts.

Weakly supervised learning is a machine learning paradigm to deal with time- and cost-sensitive annotation of training data by learning predictive models utilizing inaccurate or noisy labels. We use Snorkel~\cite{ratner-vldb17} for programmatically creating training data with weak supervision. After generating weak labels with Snorkel, we randomly sample 20K tweets with label confidence scores greater than 80\% from each of $D_J$ and $D_C$. The sampled data and sample weights based on confidence scores are used to train a Gradient Boosting Machine (GBM) classifier from scikit-learn library.

Snorkel allows users to write arbitrary labeling functions that label data points or abstain from labeling. The labeling functions are soft rules and have unknown accuracies, correlations, or conflicts. The outputs of the labeling functions are automatically modeled by a generative model that produces probabilistic labels that can be used to train a discriminative model for classification. We define four labeling functions with three functions applying heuristics for capturing personal tweets and a function using transfer learning to decide whether a tweet is a personal experience:
\begin{enumerate}
    \item \texttt{if} tweet contains URL \texttt{return} ``non-personal" \texttt{else} \texttt{return} ``personal"
    \item  \texttt{if} tweet contains first-person pronouns \texttt{return} ``personal" \texttt{else} \texttt{return} ``abstain"
    \item \texttt{if} $\texttt{subjectivity}(\text{tweet})<1$ \texttt{return} ``non-personal" \texttt{else if} $\texttt{subjectivity}(\text{tweet}) > 4$ \texttt{return} ``personal" \texttt{else} \texttt{return} ``abstain"
    \item \texttt{if} $\texttt{confidence\_score}(\text{tweet})>0.6$ \texttt{return} ``personal" \texttt{else if} $\texttt{confidence\_score}(\text{tweet})<0.1$ \texttt{return} ``non-personal" \texttt{else} \texttt{return} ``abstain"
\end{enumerate}

The first labeling function is motivated by the fact that promotional and news-related tweets usually contain an URL. The second labeling function is based on the assumption that tweets with first-person pronouns are more likely to be personal tweets than others. The third labeling function checks the subjectivity score obtained from the MPQA~\cite{wilson-emnlp05} subjectivity lexicon to determine the label of the tweets. A highly subjective tweet is likely to express personal opinion whereas a tweet with low subjectivity may be a non-personal tweet. The fourth labeling function uses the confidence scores of a tweet being a personal experience obtained from a transfer learning approach which we introduce next. The thresholds are chosen to encourage low false positives.

A transfer learning approach focuses on utilizing knowledge gained by a model in a source domain to a destination domain. In our case the destination domains are cannabis and e-cigarettes.
We chose two source domains, dietary supplement use~\cite{jiang-aclworkshop16} and medication use~\cite{jiang-bmc18} for the task of personal experience classification. First, we apply instance-based domain adaptation~\cite{pan-kde09} to select $5.7$K samples from the source datasets\footnote{https://github.com/medeffects/tweet\_corpora/} that are the most similar to the sample of $50$K tweets each from $D_J$ and $D_C$ target dataset. The cosine similarity of word vectors obtained from pre-trained Glove twitter word embeddings~\cite{pennington-emnlp14} is used as the distance metric to choose source instances that are the closest to the target instances. Next, a deep learning-based classifier similar to one employed by~\citet{jiang-bmc18} is used to obtain a confidence scores on a tweet expressing personal experiences.

We evaluate the personal tweet classification pipeline by annotating $500$ tweets from $D_J$ that are randomly sampled from the top $1,500$ tweets categorized as personal experiences by the transfer learning classifier. We also annotate $500$ tweets from $D_C$ using the same process. The tweets are annotated by trained graduate students where each tweet is labeled by two annotators. The graduate students are asked to answer the multiple-choice question ``To the best of your judgment, is this tweet about a personal experience, opinion, or observation?" with one of the options ``Yes," ``No," or ``Not sure."
The observed average agreement between the annotators is $94.6\%$ with Krippendorff's alpha coefficient equal to $0.47$ for e-cigarettes tweets. For cannabis tweets, the inter-annotator observed agreement is $85.6\%$ with Krippendorff's alpha coefficient of $0.55$. We only use the labels with an agreement between the annotators for the evaluation of the pipeline. The macro averaged F1-score of the e-cigarette evaluation dataset increases from $49\%$ by the transfer learning classifier to $71\%$ by Snorkel's generative model to $94\%$ by the discriminative GBM classifier. Similarly, the macro averaged F1-score of the cannabis evaluation dataset changes from $47\%$ by transfer learning to $89\%$ by Snorkel to $87\%$ by GBM classifier. We choose confidence score thresholds of $0.1$ and $0.5$, based on performance in the evaluation dataset, for categorizing tweets in $D_J$ and $D_C$ as personal tweets respectively.  $83\%$ of tweets in $D_J$ are categorized as personal tweets whereas only $45\%$ of tweets in $D_C$ fall into personal tweets with the majority of cannabis-related tweets being promotional or related to news. We retain the tweets from user accounts that have majority of personal tweets for further analysis.

We check for the presence of bots for the user accounts selected for further analysis. We use seven available lists of bot accounts previously used by other studies~\cite{broniatowski-ajph18,tian-aaai20} to filter out bot accounts in our dataset. We observe that $33$ out of $37$ bot accounts present in the original dataset are filtered out by the personal tweet selection pipeline demonstrating some robustness toward automated accounts.
Interestingly, we notice the remaining four bot accounts, which we filter out from further analysis, produce more human-like content.
Let $M_{JP}$ and $M_{CP}$ respectively be the retained tweets from $D_J$ and $D_C$.

\begin{table*}[!t]
    \centering
    \begin{tabular}{|p{0.25\textwidth}|>{\raggedright\arraybackslash}p{0.7\textwidth}|}
    \hline
        \textbf{Topic Label (Dataset)} & \textbf{Tokens ranked by frequency (with minimum 1\% support)} \\
        \hline
        Nicotine addiction($M_{JP}$) & nicotine\_addiction, dollar, addiction, generation, cool\_be, puff, cigarette\_pack, think\_juul, s\_juul, juul\_addiction, never\_smoke, air, have\_time, fight, have\_dollar, iphone \\
       \hline
       Smoking ($M_{JP}$) & smoke, smoke\_juul, quit, feel, smoke\_cigarette, rn, smoke\_cigs, not\_do, quit\_smoke, cloud, fun, bad\_be, stop\_smoke, everybody, quit\_juul, fool, smoke\_weed \\
        \hline
        
         Teenagers and parents ($M_{JP}$) & get\_juul, mom, find\_juul, dad, end, leave\_juul, when\_hit, find\_mom
\\
\hline

Purchase experience ($M_{JP}$) & 
pod, juul\_pod, buy, buy\_juul, buy\_pod, be\_pod, pod\_pack, mango\_juul\_pod, get\_pod\\
       \hline
Flavors and accessories ($M_{JP}$) & thing, shit, room, weekend, juul\_room, juul\_shit, cucumber\_pod, order, juul\_thing, mean, pen, only\_thing, juul\_cucumber\_pod, lose\_juul\_charger, first\_thing, fuckin\_juul, customer \\
       \hline
Cannabis odor ($M_{JC}$) & shit, smell, smell\_weed, room, weed\_amp, head, walk, somebody, pull, only\_thing, want\_smoke \\
       \hline
Cannabis legalization ($M_{JC}$) & marijuana\_legalization, support, legalization, company, marijuana\_cannabis, ballot\\      
\hline
Cannabis consumption ($M_{JC}$) & way, thc, dog, good\_be, bud, cannabis\_industry, marijuana\_plant, lung, kick, flower, weed\_be, marijuana\_bill, wax, hash\_oil, try\_weed \\
\hline

    \end{tabular}
    \caption{Relevant topics and concept tokens ranked by frequency.}
    \label{tab:lda}
\end{table*}

\subsection{Topic Analysis}
We perform topic analysis on the tweets retained by the personal tweet selection approach. We use Latent Dirichlet Allocation (LDA) available via python's scikit-learn library for automatically extracting topics. To obtain more meaningful and interpretable topics, we run LDA using concept tokens extracted from tweets instead of generally used word tokens or word n-grams. We extend the graph-based concept extraction technique \cite{rajagopal-www13} to handle text data from tweets. First, we use TweeboParser\footnote{https://github.com/ikekonglp/TweeboParser} \cite{kong-emnlp14} library to tokenize tweets, assign the tokens with Part-of-Speech(POS) tag, and construct a dependency tree. Next, we utilize the dependency structure and POS tags to extract concepts combining lemmatized tokens. We capture linguistic features such as noun phrases, verb phrases, and adjectival and adverbial modifications. 
We extract $20$ topics each from the tweets in two data sets $M_{CP}$, and $M_{JP}$ independently. Using other default parameters of scikit-learn LDA, we treat the number of topics as a hyperparameter and use the perplexity score metric as well as manual inspection for selecting $20$ topics. The topic analysis confirms that users are sharing their opinions or experiences related to JUUL and cannabis use. Table \ref{tab:lda} shows example topics with top-weighted concepts automatically extracted by LDA. The users in $M_{JP}$ are discussing opinions on nicotine addiction, experiences of smoking, purchase of refill for e-cigarette device, JUUL flavors and accessories, and JUUL experiences together with parent encounters.
The individuals in $M_{CP}$ discuss cannabis legalization policies, the encounter of cannabis odor, and modes of cannabis consumption. The last row also captures the topics suggesting dual-use. The e-cigarette device may be used to intake cannabis in the form of dried ``buds," ``wax,"  or ``THC."

\subsection{Stance Detection} 
\label{sec:st_dec}
For the stance detection task, we first determine the stance of each personal tweet by classifying the tweet into one of "In favor," "Against," or "Neither" classes. In this section, we describe the annotation procedure and classification methodology. 
\subsubsection{Annotation.} 
We perform annotation of $3,000$ tweets to prepare a dataset\footnote{\label{repo}https://github.com/edgeslab/vaping} for training and evaluation. We select the top $1,500$ tweets from each of $D_J$ and $D_C$ with the highest confidence from the transfer learning (TL) classifier. For each $1,500$ tweets, we sample $500$ tweets for annotation by trained graduate students and $1,000$ tweets for annotation by Amazon Mechanical Turk (MTurk) workers. We ask two multiple-choice questions for each tweet. The questions asked for sampled cannabis tweets are as follows: \\
Q1. ``To the best of your judgment, is this tweet referring to cannabis?" Answer: Yes or No.\\
Q2. ``To the best of your judgment, is the person who wrote this tweet in favor of or against cannabis use?" Answer: In favor, Against, or Neither. \\
The e-cigarette annotation included the same questions where ``cannabis" was replaced with ``e-cigarettes." The answers to the first question helped us evaluate the tweets filtering process using keywords and hashtags by accessing the fraction of false positives for the tweets selected for annotation.
The answers to the second question help us prepare a dataset to train and evaluate the tweet stance classification model.

We use Labelbox\footnote{https://labelbox.com/} for internal annotation and training of graduate students. For internal annotation, each tweet is annotated by two graduate students. The workers for crowdsourcing were selected such that they are from the United States, have an overall approval rate of greater than $95\%$, and have worked on at least $100$ tasks. In addition to the worker qualifications, we add a quiz question as an additional quality assurance step. The quiz question includes a random example from the instructions and the workers need to answer it correctly for their work to be accepted. The detailed instructions provided to crowdsourced workers are available as supplemental $\text{material}^\text{\ref{repo}}$. 
Each worker is asked to answer two questions for each of five tweets. A task is assigned to three workers and labels are decided by majority voting. If there is no majority on the second question, the conflict is resolved by an additional annotation by a graduate student. Table \ref{tab:iaa} summarizes the observed average inter-annotator agreement (IAA) percentages and Krippendorff's alpha coefficients for internal and crowdsourced annotators.

\begin{table}[!t]
    \centering
    \begin{tabular}{c c c c c}
    \hline
    Q.No. & Platform & Dataset & IAA \% & $K_{\alpha}$ \\
    \hline
         Q1 & Labelbox & JUUL & $98.6$ & $-0.006$ \\
         Q1 & Labelbox & Cannabis & $96.0$ & $0.52$ \\
         Q1 & MTurk & JUUL & $99.5$ & $0.61$ \\
         Q1 & MTurk & Cannabis & $97.1$ & $0.63$ \\
         Q2 & Labelbox & JUUL & $80.4$ & $0.55$\\
         Q2 & Labelbox & Cannabis &  $71.0$ & $0.46$\\
         Q2 & MTurk & JUUL &  $81.7$ & $0.58$\\
         Q2 & MTurk & Cannabis & $68.6$ & $0.43$\\
         \hline
    \end{tabular}
    \caption{Observed inter-annotator agreement (IAA) and Krippendorff's alpha ($K_{\alpha}$) scores for annotated samples described in the section on Stance Detection.}

    \label{tab:iaa}
\end{table}
\begin{table}[!t]
    \centering
    \begin{tabular}{|c|c|c|c|}
    \hline
        Dataset $(\downarrow)$/Labels$(\rightarrow)$ & In favor & Neither & Against \\
        \hline
         E-cigarettes & $76\%$ & $17\%$ & $7\%$\\
         Cannabis & $54\%$ & $42\%$ & $4\%$\\
         \hline
    \end{tabular}
    \caption{Stance distribution in the annotated tweets.}
    \label{tab:stImb}
\end{table}

\subsubsection{Tweet stance classification.}
We learn two classifiers separately for each of the datasets $M_{JP}$ and $M_{CP}$ using the tweet word vector features. We use the crowdsourced annotations for training and internal annotations for evaluation. The dataset for tweet level stance classification obtained after annotation shows imbalanced distribution with more tweets in favor of e-cigarette and cannabis use as shown in Table \ref{tab:stImb}.
We perform model selection and hyperparameter tuning to choose a stance classification model robust to class imbalance and overfitting. Moreover, we want the classifier to have a low categorical cross-entropy for the predictions to ensure a less noisy probabilistic classifier. Here, we report the best performing classifier after experimenting with logistic regression, support vector machine (SVM), and an LSTM-based deep learning model for stance classification. Logistic regression with balanced class weights and complexity parameter $C=1$ achieves the best weighted AUC of $81.0\%$ (micro-AUC $92.9\%$) for e-cigarette and $75.4\%$ weighted AUC (micro-AUC $85.3\%$) for cannabis evaluation data. Additionally, we use the scikit-learn library's probability calibration to improve the cross-entropy loss from $0.74$ to $0.48$ for e-cigarette and $0.82$ to $0.71$ for cannabis evaluation data, performed with $20\%$ held out data. The predicted class probabilities are added to the tweet instances in $M_{JP}$ and $M_{CP}$ as a derived attribute $m_i.p_s$.

\subsubsection{User stance detection.}
Here, we describe our approach to detect a user's overall stance by aggregating stance expressed in their tweets for a given period. The labels "In favor," "Neutral, " and "Against" obtained after probabilistic classification are assigned polarity scores of $1$, $0$, and $-1$, respectively. The overall stance for a user in a given period is the sign obtained after adding all the polarity scores. A user is defined as pro-JUUL if they have an overall positive polarity score based on tweets before a treatment state's legalization date. A user is said to be pro-cannabis if they have an overall positive polarity score based on tweets after a treatment state's legalization date and up to the end of the study period. 
An alternative user stance aggregation approach would be to give more weight to recent tweets.

\section{Causal Effect Estimation}

Next, we describe how we estimate the causal effect of recreational cannabis legalization on developing a pro-cannabis attitude for e-cigarette users, a proxy for escalation from tobacco e-cigarette use to cannabis use, for three U.S. states. We consider a number of possible model choices, including type of matching and propensity score classifier, in order to perform sensitivity analysis and determine which model produces the best covariate balance and thus less biased causal effect estimation. 

\textbf{Population of interest.}  
Our population of interest is pro-JUUL users from the United States with no cannabis-related tweets before the recreational legalization effective date of the treatment state.
During our study period, from January 2016 to December 2018, six U.S. states, namely Nevada, Maine, Massachusetts, California, Vermont, and Michigan, have enforced recreational legalization of cannabis. We exclude Nevada and Maine from the analysis due to the low number of users in the treatment group. We also exclude Michigan as we have an outcome only for the first month after its legalization effective date. We focus on the remaining three, California (CA), Massachusetts (MA), and Vermont (VT), and consider each one individually as a treatment state. Users from all other 49 US states and DC are considered for the control group.

\textbf{Treatment/Control assignment.} We assign users from a ``treatment" state $S$ with newly legalized recreational cannabis to the treatment group ($T^{(S)}$). We assign other users to four control groups based on the cannabis policy of their state on the legalization effective date of the treatment state: 
\begin{itemize}
    \item {\small $C_1^{(S)}$}: Users from all states with illegal cannabis
    \item {\small $C_2^{(S)}$}: Users from all states with legal medical cannabis with limited THC
    \item {\small$C_3^{(S)}$}: Users from all states with legal medical cannabis
    \item {\small $C_4^{(S)}$}: Users from all states with previously legalized recreational cannabis
\end{itemize}
The legalization effective dates for CA, MA, and VT are January 1,2018, July 28, 2017, and July 1, 2018 respectively.

To estimate a $95\%$ confidence interval of the causal estimate, we run each experiment over $200$ samples obtained by sampling tweet stance based on the stance classifier probabilities ($m_i.p_s$), determining user stance for each sample and then assigning them to treatment or control when their stance towards e-cigarettes is in favor prior to cannabis legalization. Table \ref{tab:sample_size} shows the mean and standard deviation of the population of interest belonging to treatment and control groups obtained after running $200$ simulations. The comparatively low population size is likely due to the selection criteria for the study which require that a user has expressed opinion in favor of e-cigarettes but has not mentioned cannabis prior to the treatment state's legalization date. There are comparatively few users from Vermont which is also one of the smallest states by population size. The reason behind the low sample size for Massachusetts is likely that the cannabis legalization date was prior to JUUL becoming very popular.

\begin{table}[!t]
    \centering
    \begin{tabular}{|l|c|c|c|c|c|}
        \hline
        \multirow{2}{2em}{State} & \multicolumn{5}{c|}{Sample Size} \\
        \cline{2-6}
          & $T$ & $C_1$ & $C_2$ & $C_3$ & $C_4$\\
        \hline
        \multirow{2}{2em}{CA}&{\small$126$} & {\small$511$} & {\small$2042$} & {\small$4947$} & {\small$1148$} \\
         & {\small$(27)$} & {\small$(15)$} & {\small$(30)$} & {\small$(63)$} & {\small$(23)$}\\
         \hline
        \multirow{2}{2em}{MA}&{\small$37$} & {\small$33$} & {\small$220$} & {\small$997$} & {\small$113$} \\
         & {\small$(4)$} & {\small$(4)$} & {\small$(9)$} & {\small$(20)$} & {\small$(6)$}\\
      \hline
       \multirow{2}{2em}{VT}&{\small$49$} & {\small$1476$} & {\small$7729$} & {\small$13709$} & {\small$6248$} \\
         & {\small$(3)$} & {\small$(25)$} & {\small$(60)$} & {\small$(115)$} & {\small$(71)$}\\
         \hline
    \end{tabular}
    \caption{Mean (and standard deviation) of sample sizes (rounded) for treatment and control groups obtained from probabilistic stance classifier with $200$ simulations.}

    \label{tab:sample_size}
\end{table}

\textbf{Observed outcome.} We define a binary outcome $Y$ that indicates whether a user initiated a pro-cannabis tweet within $N\in[1,6]$ months of the legalization effective date of the treatment state.

\textbf{Selecting covariates.} We use the pre-treatment tweets of users represented as a mean of $n$-dimensional word embeddings as the potential confounding covariates. We treat the dimension of word embeddings as a hyperparameter that tunes covariates balance based on sample size.

\textbf{Matching models.}
We use three matching methods: 1:1 nearest neighbor matching (NNM), 1:1 propensity score matching (PSM), and normalized inverse probability of treatment weighting (IPTW). We also compare two propensity score classifiers, logistic regression (LR) and Gradient Boosting Machine (GBM), used by PSM and IPTW. 
We present the results for IPTW with Logistic Regression (IPTW-LR) which produces the most balanced samples. IPTW is also the recommended matching method for ATE estimation according to ~\citet{stuart-sc10}. We also discuss the results of covariate balance and causal effect estimation by IPTW-GBM, PSM-LR, PSM-GBM, and NNM, together with IPTW-LR without considering retweets.

 We use the nearest neighbor matching with replacement~\cite{stuart-sc10} approach for both NNM and PSM models. For NNM, we match pre-treatment covariates using cosine similarity-based distance metric as follows:
 {\small
\begin{equation}\label{eq:cos_dis}
    D_{cosine}(v_i.X, v_j.X) = 1 - \frac{v_i.X \cdot v_j.X}{\norm{v_i.X} \times \norm{v_j.X}}
\end{equation}
}
For PSM, we match the propensity scores with a distance metric of squared Euclidean distance, {\small $D_{ps}(v_i.e, v_j.e) = (v_i.e - v_j.e)^2$}.
The propensity score model predicts likelihood of treatment, i.e., belonging to a state with newly legalized cannabis given the covariates.
We handle the class imbalance in LR and GBM propensity score models to ensure the propensity score model is not biassed toward the majority class. Since we have multiple control groups, we define one propensity score model for each control group resulting in four models. 
For IPTW, the sample weight for user $v_i$ is determined as
    {\small $v_i.w = (v_i.e)^{-1}v_i.T + (1-v_i.e)^{-1}(1-v_i.T)$}. 
To avoid dividing by $0$ or by very small numbers, only scores between $0.05$ and $0.95$ are included. 

\textbf{Assessing covariate balance.}  We measure the absolute standardized mean difference (ASMD) of the distribution of the raw covariates in the treatment and control groups. The same process is repeated for samples after matching to assess whether matching produced better covariate balance in the treatment and control groups. It is expected that the value of ASMD should be a low value (typically less than $0.1$).

\textbf{ATE estimation.} The average treatment effect for NNM and PSM is computed using the balanced dataset $(\mathbf{B})$ obtained after 1:1 matching. Let $v_i$ and $v_i^M$ respectively be an instance in the treatment group and its matched instance from the control group. The ATE is given by:
{\small
\begin{equation}\label{eq:ateMatching}
    ATE_{matching} = \frac{1}{N_T}\sum_{v_i|v_i.T=1 \wedge v_i \in \mathbf{B}}{v_i.Y(1) - v_i^M.Y(0)}
\end{equation}
}
where $N_T$ is the instances in the treatment group.
The average treatment effect with IPTW is calculated using equation~\ref{eq:ate}.
{\small
\begin{equation}\label{eq:ate}
    \begin{split}
            ATE_{IPTW} = \frac{\sum_i{(v_i.e)^{-1} \times v_i.T \times v_i.Y(1)}} {\sum_i{(v_i.e)^{-1} \times v_i.T }} - \\
        \frac{\sum_i{(1-v_i.e)^{-1} \times (1-v_i.T) \times v_i.Y(0)}}{\sum_i{(1-v_i.e)^{-1} \times (1-v_i.T)}}\
    \end{split}
\end{equation}
}
where $v_i.Y(1)$ is the observed outcome for the treated individual $v_i$ and $v_j.Y(0)$ is the observed outcome for the individual $v_j$ in control group.
The $95\%$ confidence interval is calculated as
{ \small
    $E_{ci} = <E_s - 1.96 \times \sigma_b/N_b, E_s + 1.96 \times \sigma_b/N_b>$
}
where $E_s$ is the mean estimated ATE, $\sigma_b$ the standard deviation of ATE, and $N_b$ the number of simulations.

% IPTW-LR covariate balance
\begin{figure*}[!t]
\centering
    \begin{subfigure}[t]{0.32\textwidth}
         \centering
         \includegraphics[width=\textwidth]{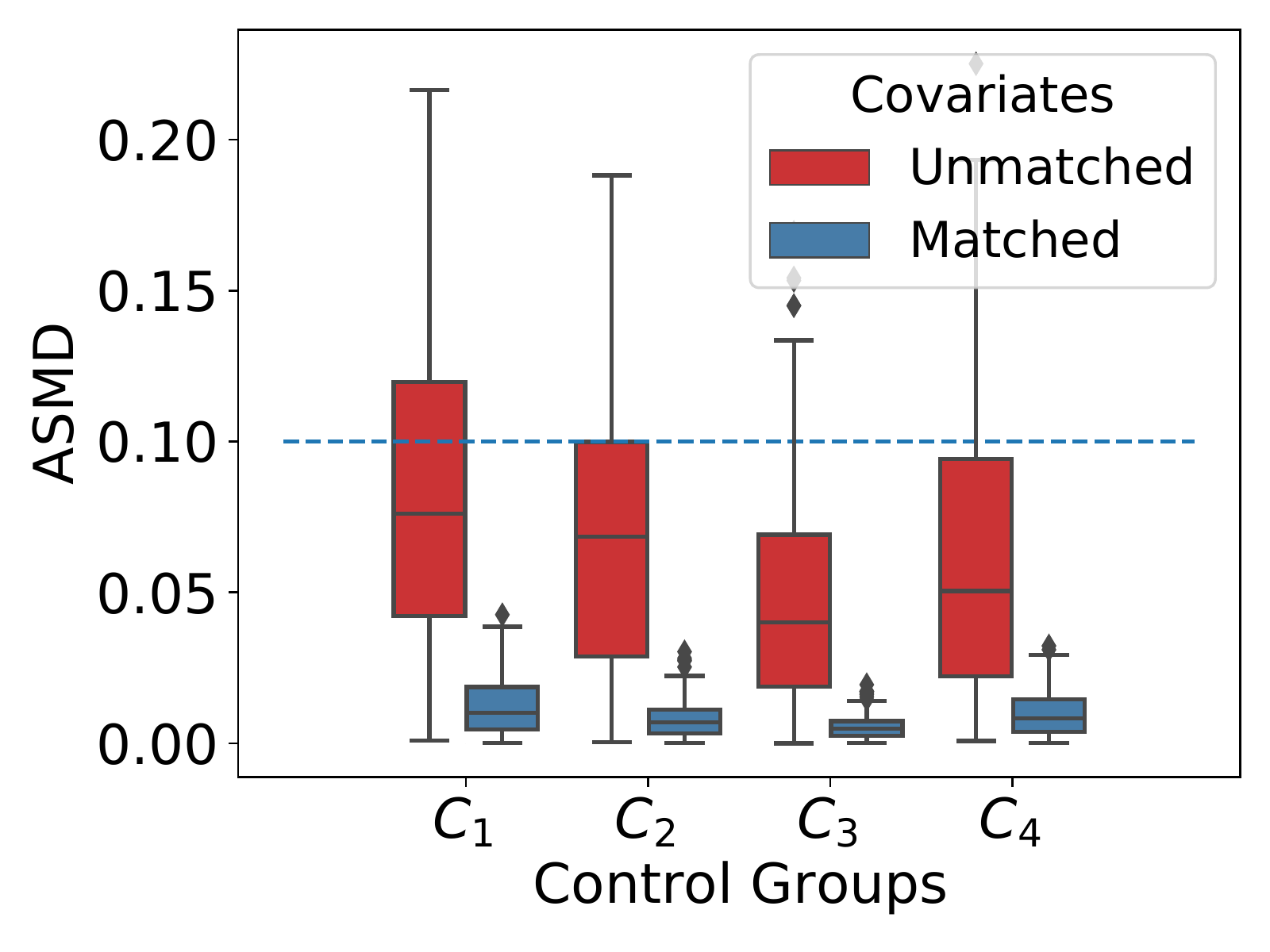}
         \caption{California}
         \label{fig:CA_bal}
     \end{subfigure}
     \hfill
     \begin{subfigure}[t]{0.32\textwidth}
         \centering
         \includegraphics[width=\textwidth]{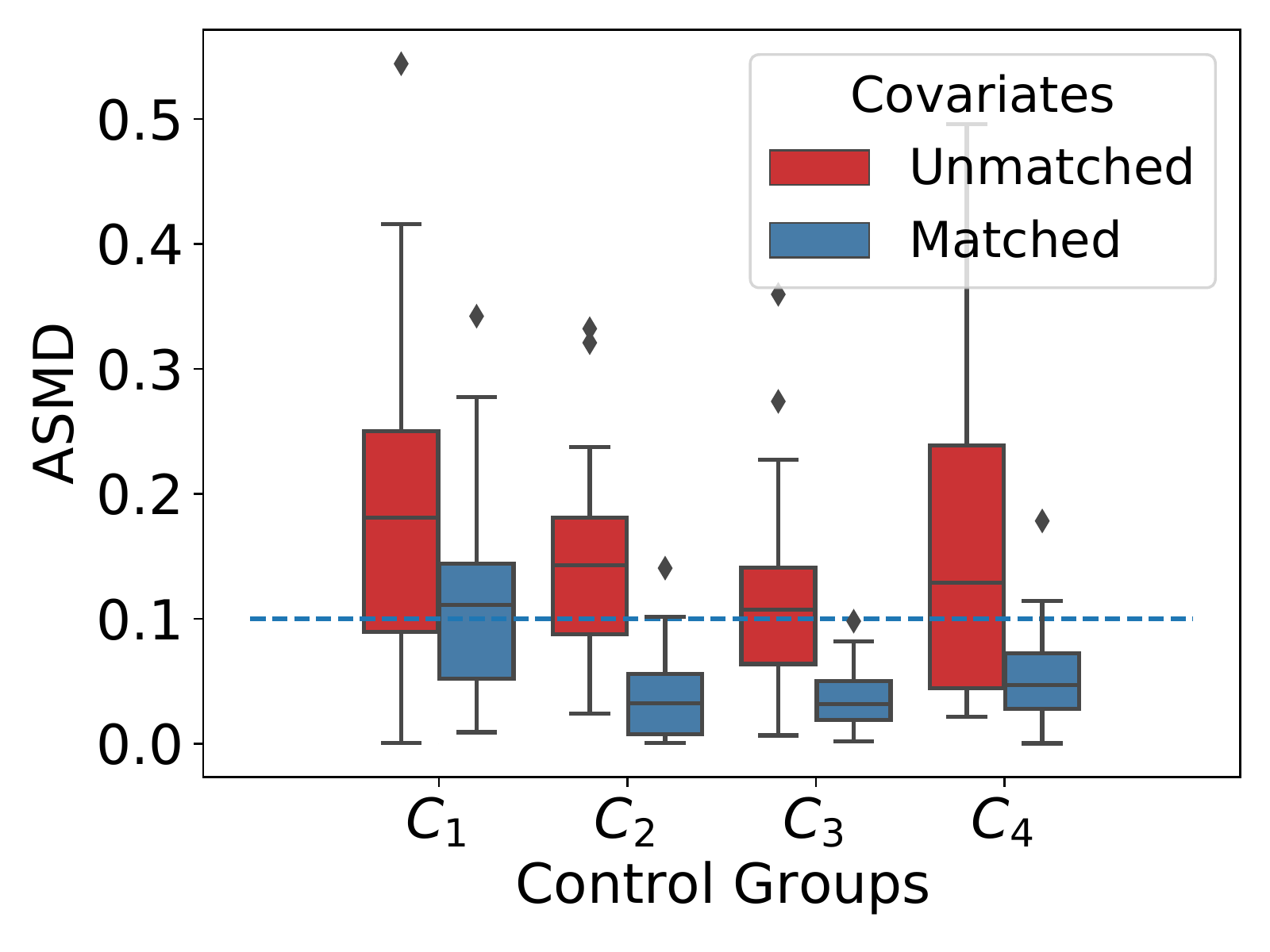}
         \caption{Massachusetts}
         \label{fig:MA_bal}
     \end{subfigure}
     \hfill
     \begin{subfigure}[t]{0.32\textwidth}
         \centering
         \includegraphics[width=\textwidth]{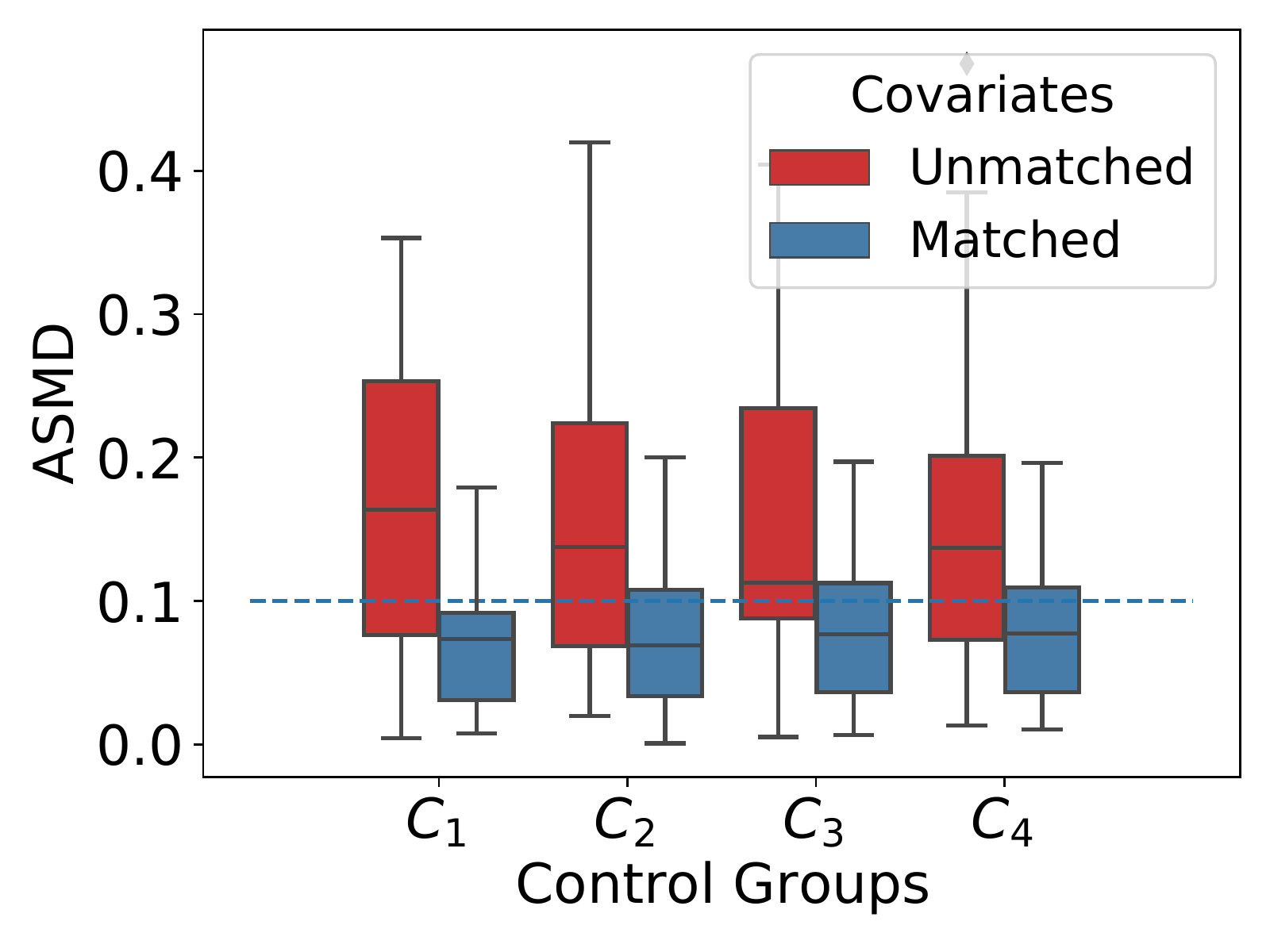}
         \caption{Vermont}
         \label{fig:VT_bal}
     \end{subfigure}
        \caption{Assessing the covariates balance after IPTW matching. Lower absolute standardized mean difference (ASMD) scores are preferred for a balanced covariate distribution in the treatment and control groups. The horizontal dashed lines indicate a threshold of 0.1 ASMD.}
        \label{fig:balance}
\end{figure*}

% IPTW-LR causal effect
\begin{figure*}[!t]
\centering
    \begin{subfigure}[t]{0.32\textwidth}
         \centering
         \includegraphics[width=\textwidth]{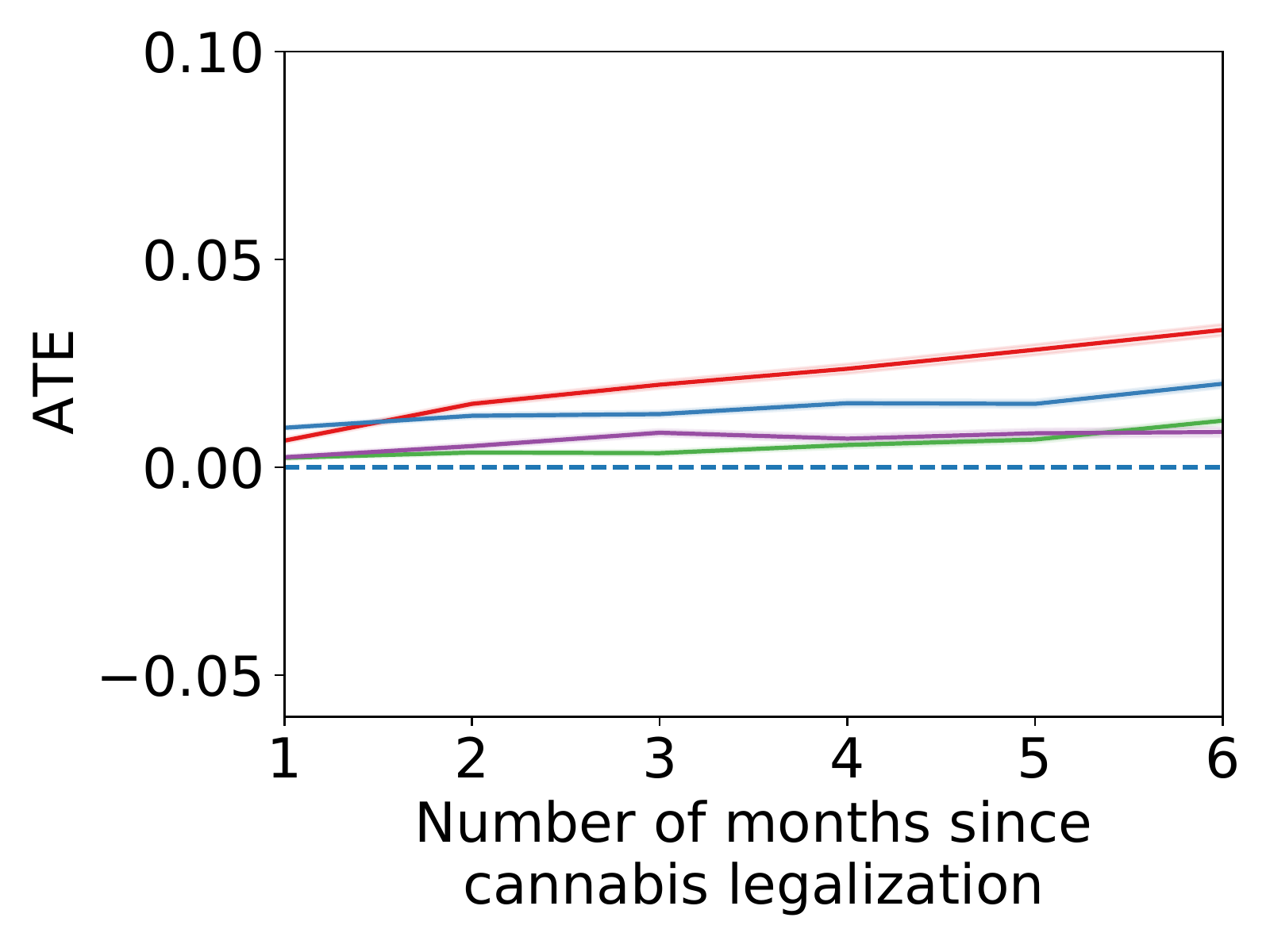}
         \caption{California}
         \label{fig:CA_eff}
     \end{subfigure}
     \hfill
     \begin{subfigure}[t]{0.32\textwidth}
         \centering
         \includegraphics[width=\textwidth]{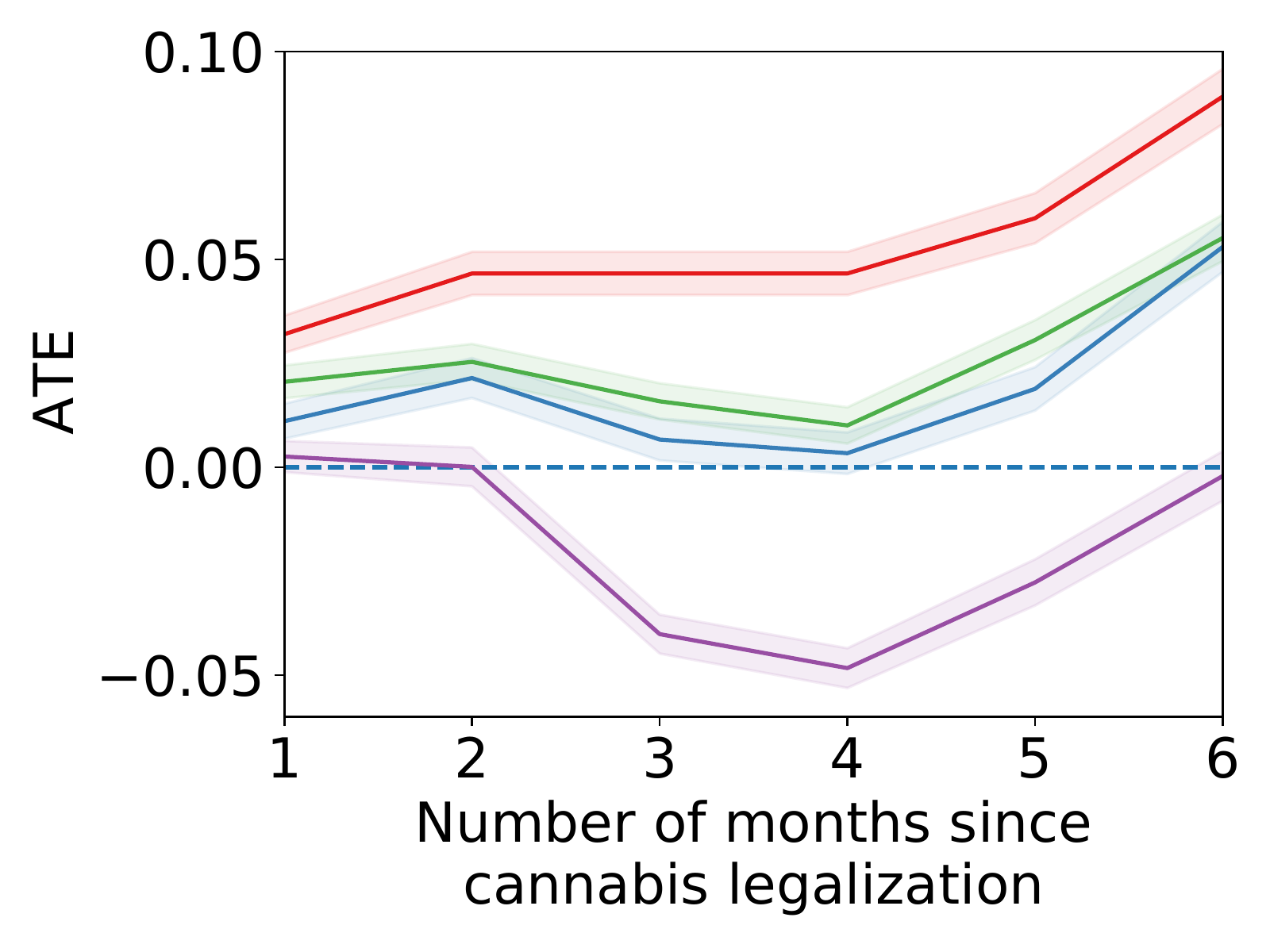}
         \caption{Massachusetts}
         \label{fig:MA_eff}
     \end{subfigure}
     \hfill
     \begin{subfigure}[t]{0.32\textwidth}
         \centering
         \includegraphics[width=\textwidth]{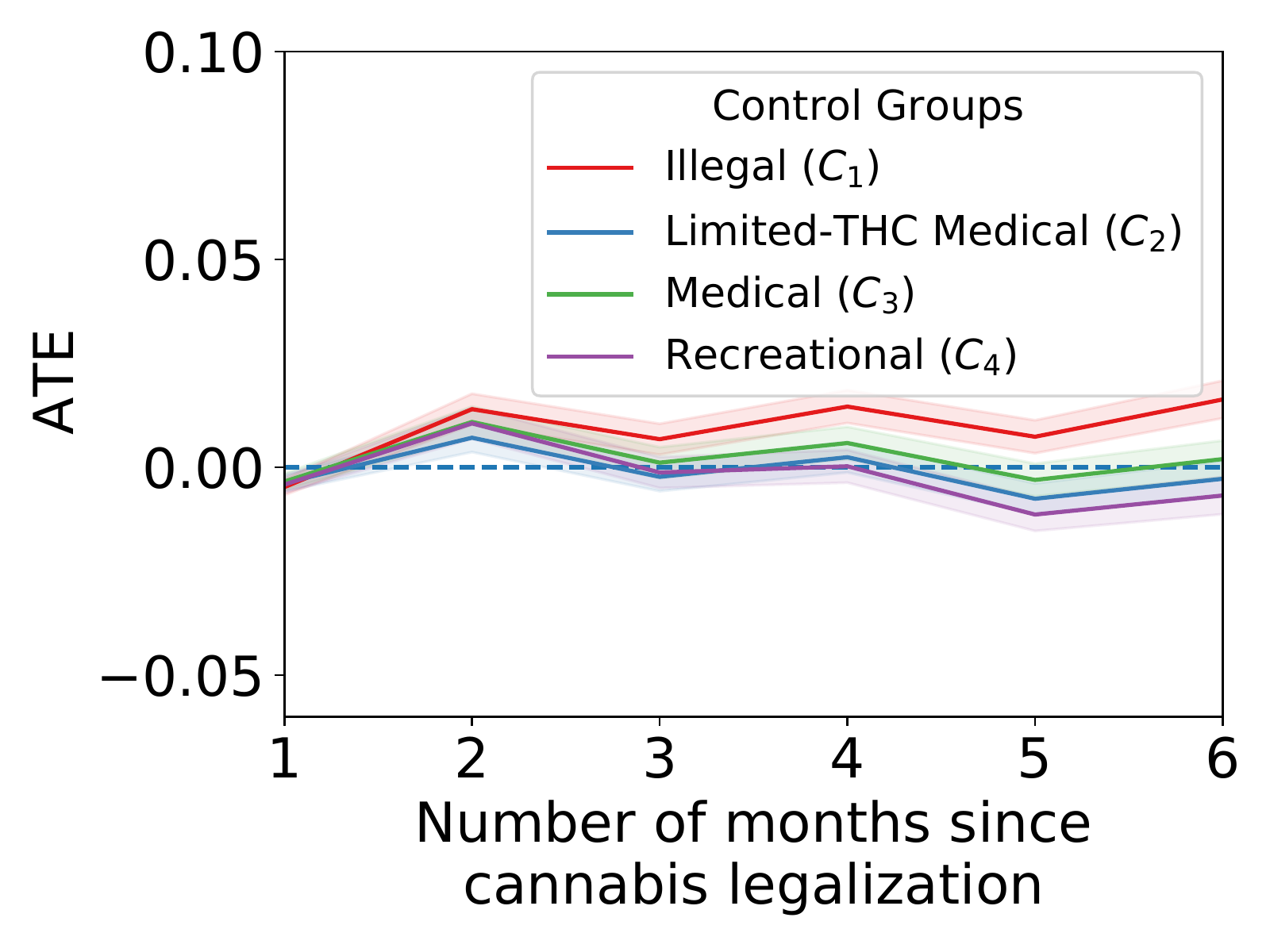}
         \caption{Vermont}
         \label{fig:VT_eff}
     \end{subfigure}
        \caption{Average Treatment Effect (ATE) of recreational cannabis legalization policy compared to control groups based on state's cannabis policy. The x-axis shows the number of months since recreational cannabis legalization in the treatment state.}
        \label{fig:effect}
\end{figure*}

\textbf{Causal effect of recreational cannabis legalization.}
Figure \ref{fig:balance} shows the box plot of ASMD scores of all covariates across the treatment group and each control group, before and after applying IPTW-LR. For California, matching results in a more balanced distribution with lower ASMD values. For Massachusetts and Vermont, although the overall covariate balance improves after matching, the improvement is not as good as that of California. Due to the relatively low average treatment population size in Massachusetts and Vermont, we use $25$-dimensional word embedding compared to $200$-dimensional in California. Additionally, California has a better covariate balance without matching compared to the other two states.

Figure \ref{fig:effect} depicts the average treatment effect (with $95\%$ confidence intervals) of recreational cannabis legalization policy on the initiation of pro-cannabis tweets for pro-e-cigarette users. The y-axis shows the ATE, the difference of the pro-cannabis tweet initiation rate between treatment and control groups, measured for $N$ months after the policy effective date. We notice two main trends across all three treatment states. First, ATE is highest and increasing over time for control groups with illegal cannabis ($C_1$). This implies that the cannabis tweet initiation rate is much higher in the states with newly legalized recreational cannabis compared to states in $C_1$. Second, ATE is close to zero or negative for control groups with already legalized recreational cannabis ($C_4$). This implies the cannabis tweet initiation rate in states with already legal recreational cannabis is similar to or higher than a state with newly legalized cannabis. The other two control groups with some form of medical cannabis legalization ($C_2$ and $C_3$) are in-between $C_1$ and $C_4$ which makes intuitive sense. There is some variance in the trend across the three treatment states for $C_2$ and $C_3$.

% Model selection covariate balance
\begin{figure}[!t]
    \centering
    \begin{subfigure}[t]{0.45\textwidth}
    \includegraphics[width=\textwidth]{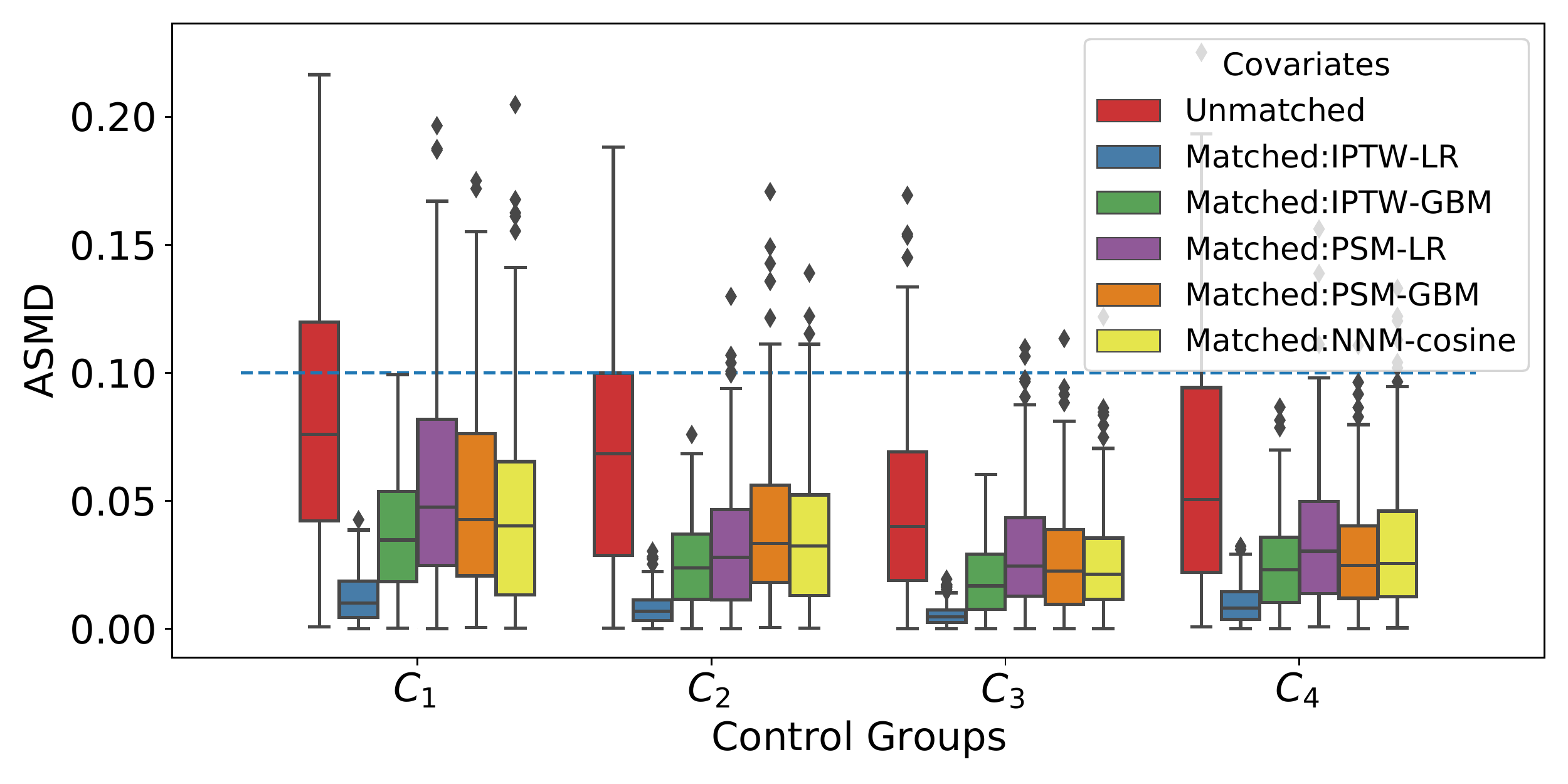}
     \caption{ASMD for matching methods when retweets are included.}
    \label{fig:sen_balance}
    \end{subfigure}
    \begin{subfigure}[t]{0.33\textwidth}
         \centering
         \includegraphics[width=\textwidth]{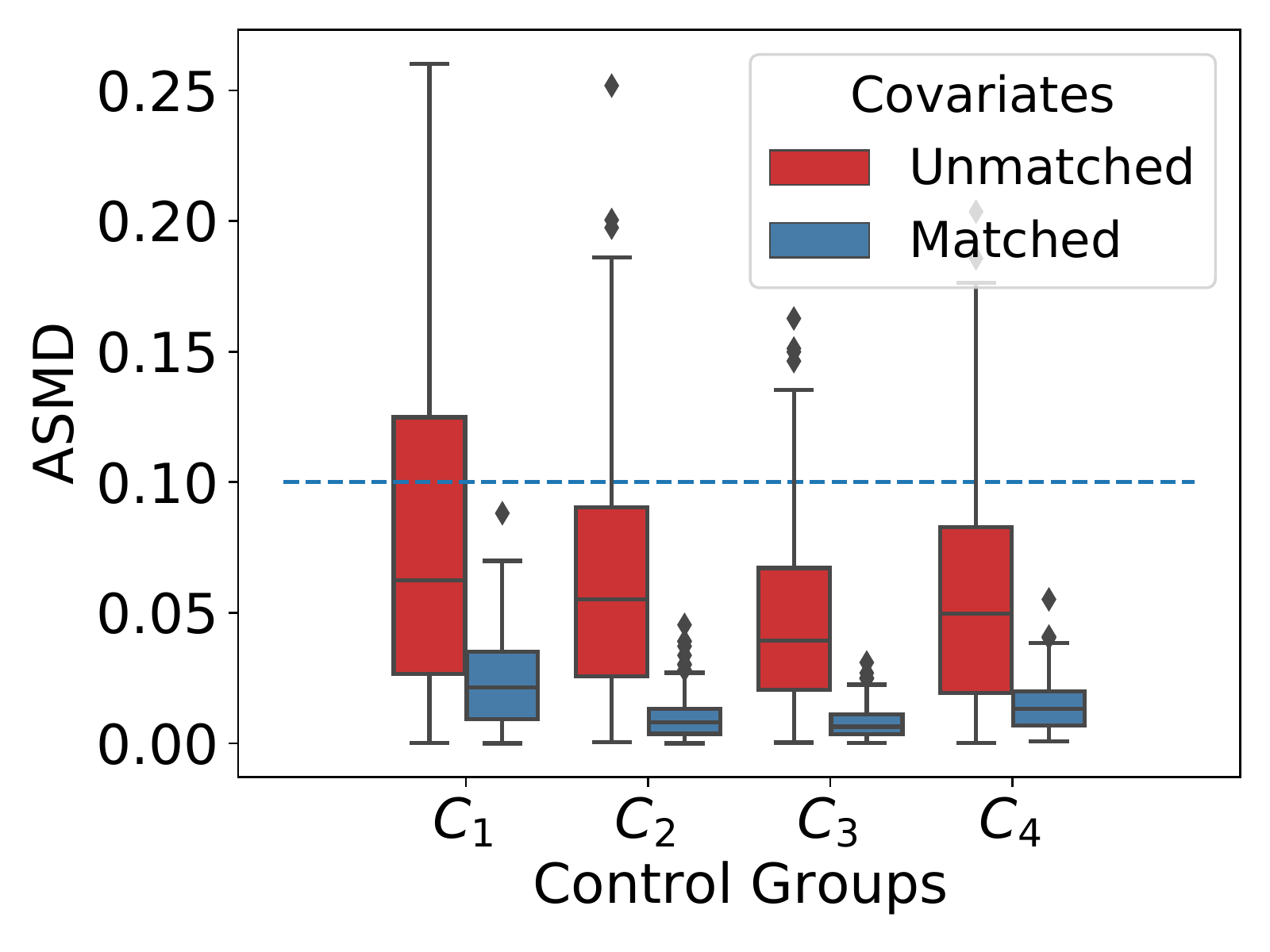}
         \caption{ASMD for IPTW with LR (best model) when retweets are excluded. }
         \label{fig:tw_iptw_lr_bal}
     \end{subfigure}
    \caption{Sensitivity analysis of covariate balance, when treatment state is California, for matching methods and inclusion/exclusion of retweets.}
    \label{fig:sen_balance_parent}
\end{figure}

% Model selection causal estimation
\begin{figure}[!t]
    \begin{subfigure}[t]{0.234\textwidth}
         \centering
         \includegraphics[width=\textwidth]{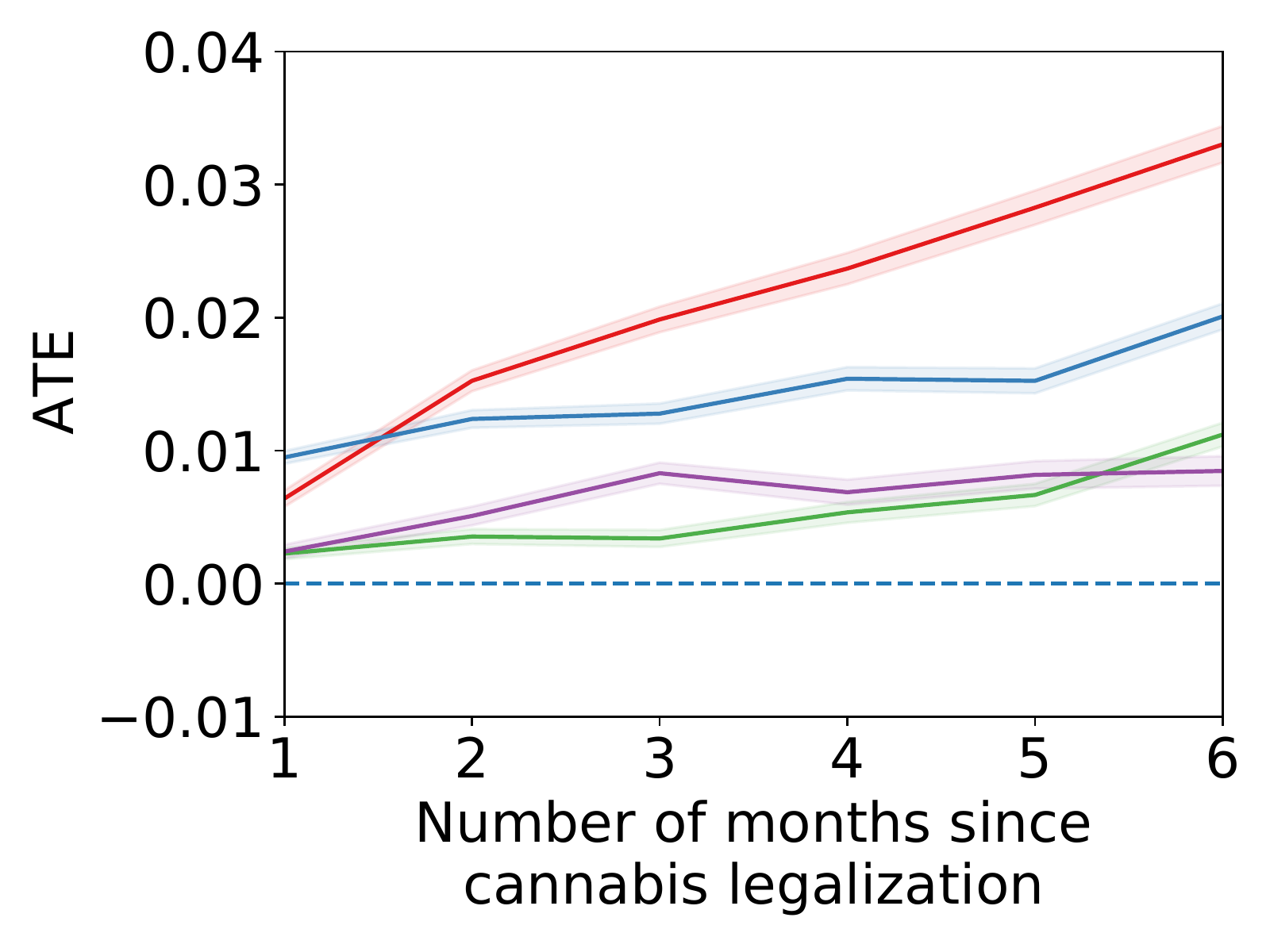}
         \caption{IPTW with LR}
         \label{fig:iptw_lr}
     \end{subfigure}
     \hfill
     \begin{subfigure}[t]{0.234\textwidth}
         \centering
         \includegraphics[width=\textwidth]{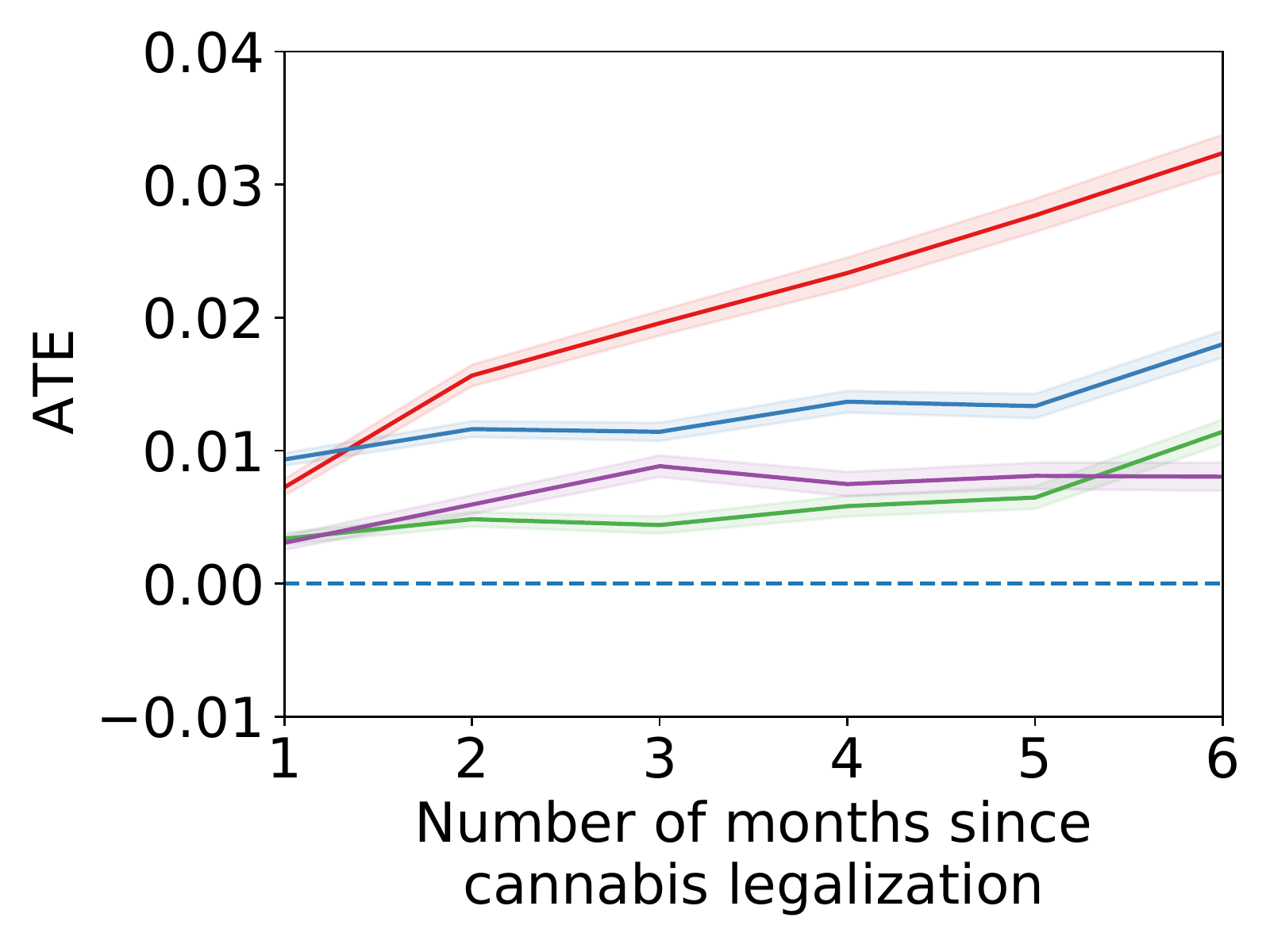}
         \caption{IPTW with GBM}
         \label{fig:iptw_gbm}
     \end{subfigure}
     \hfill
     \begin{subfigure}[t]{0.234\textwidth}
         \centering
         \includegraphics[width=\textwidth]{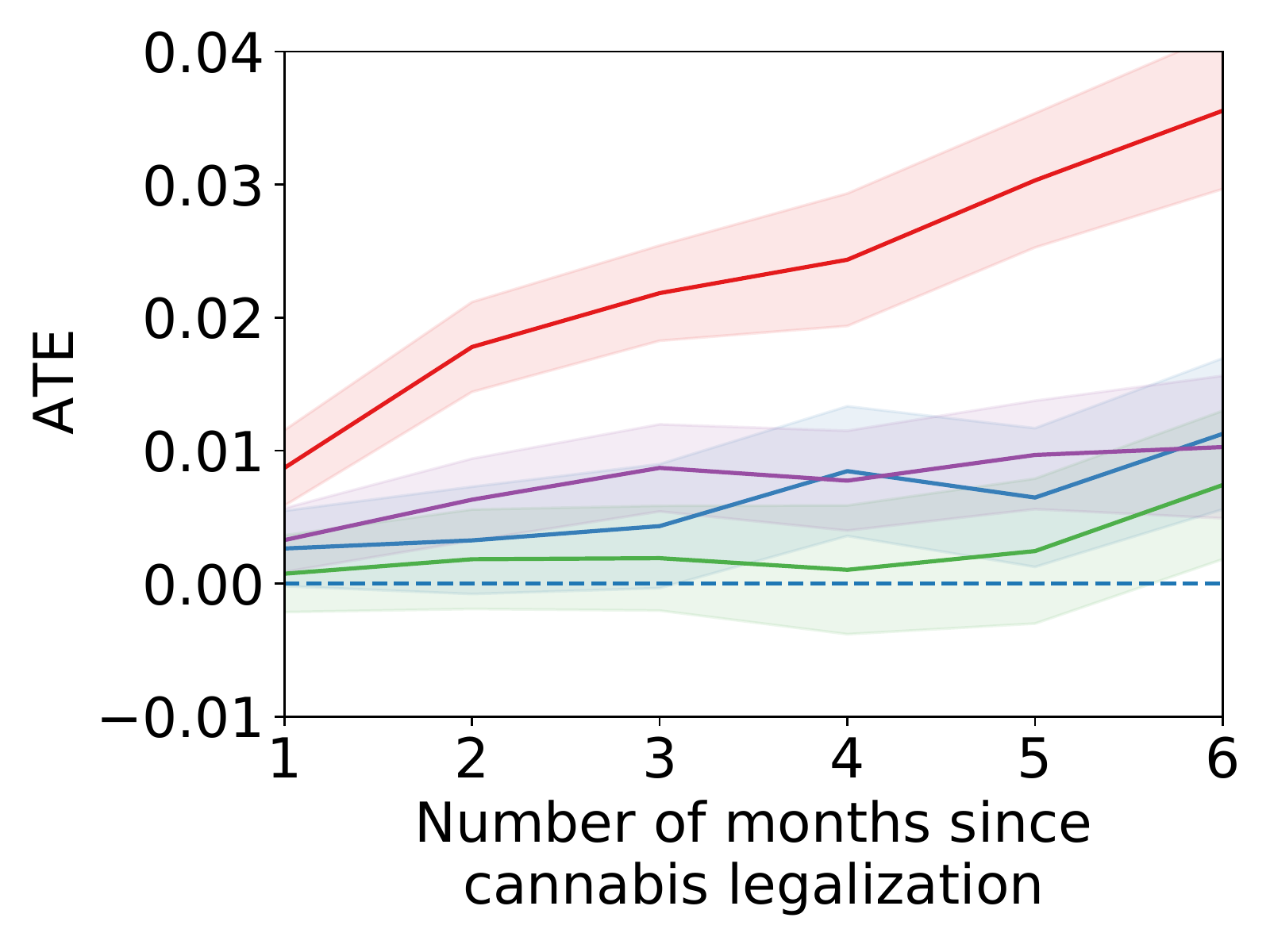}
         \caption{PSM with LR}
         \label{fig:ps_lr}
     \end{subfigure}
      \hfill
    \begin{subfigure}[t]{0.234\textwidth}
         \centering
         \includegraphics[width=\textwidth]{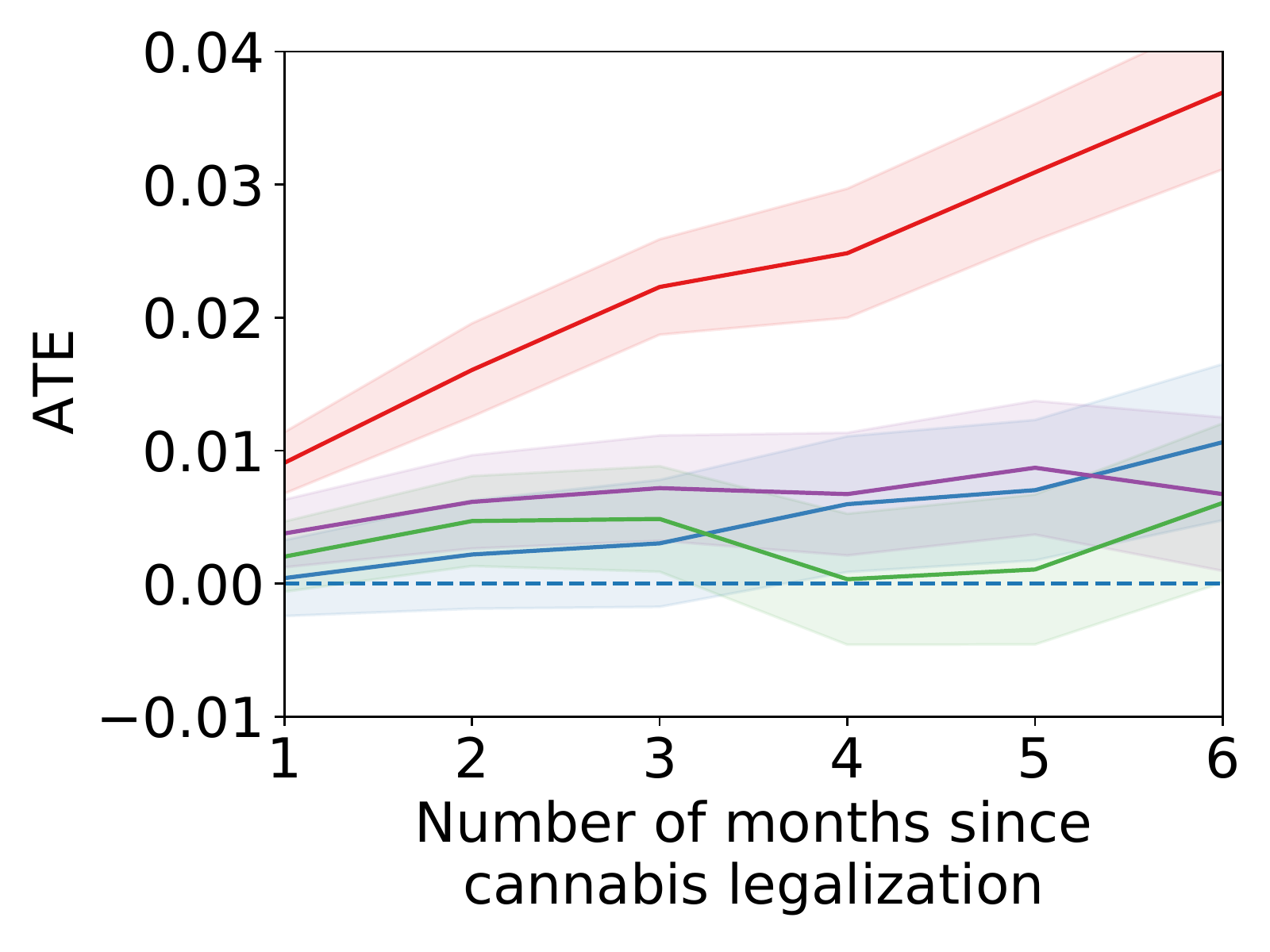}
         \caption{PSM with GBM}
         \label{fig:ps_gbm}
     \end{subfigure}
       \hfill
     \begin{subfigure}[t]{0.234\textwidth}
         \centering
         \includegraphics[width=\textwidth]{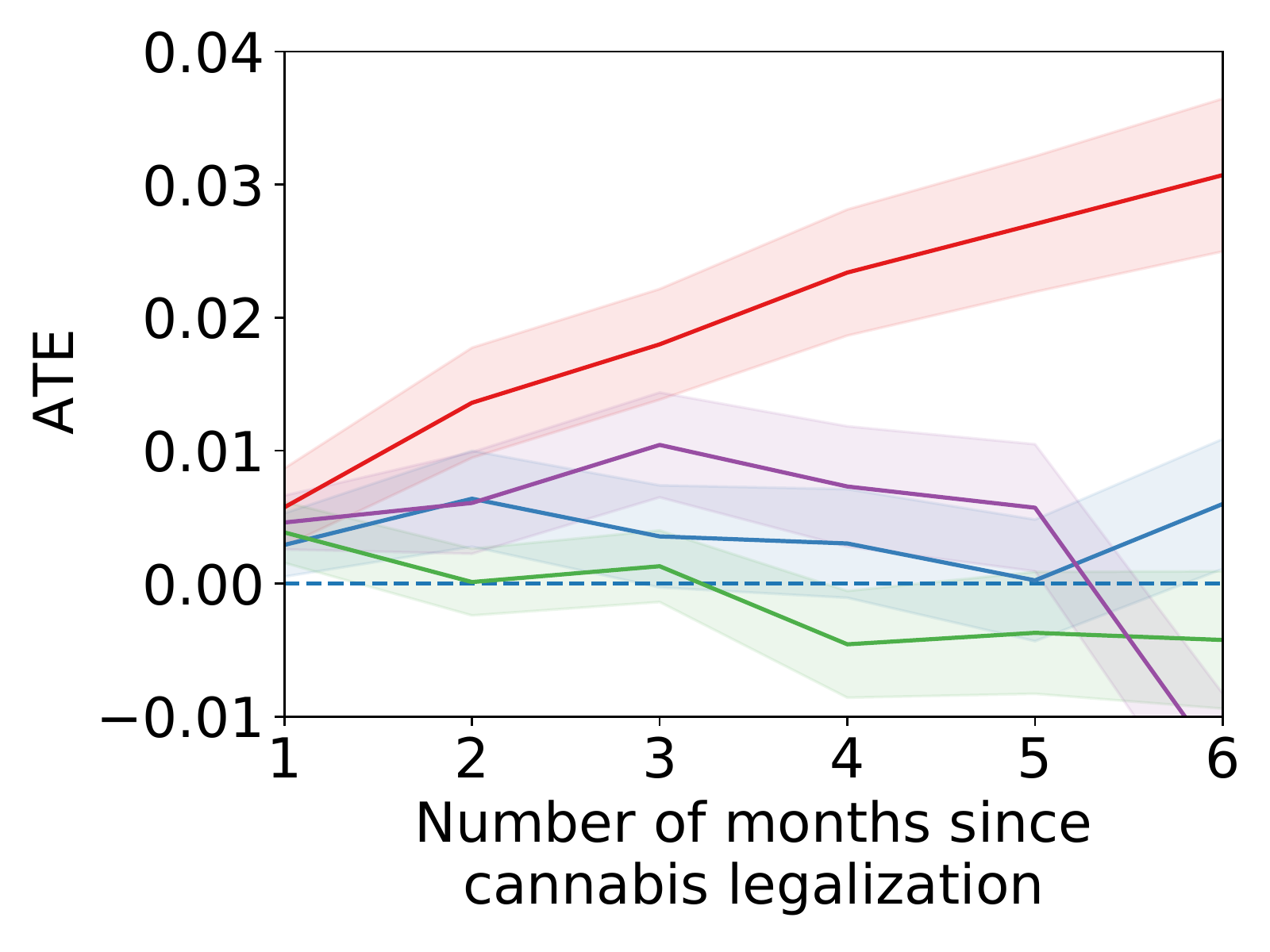}
         \caption{NNM with cosine similarity}
         \label{fig:fb_cosine}
     \end{subfigure}
    \hfill
     \begin{subfigure}[t]{0.234\textwidth}
        \centering
        \includegraphics[width=\textwidth]{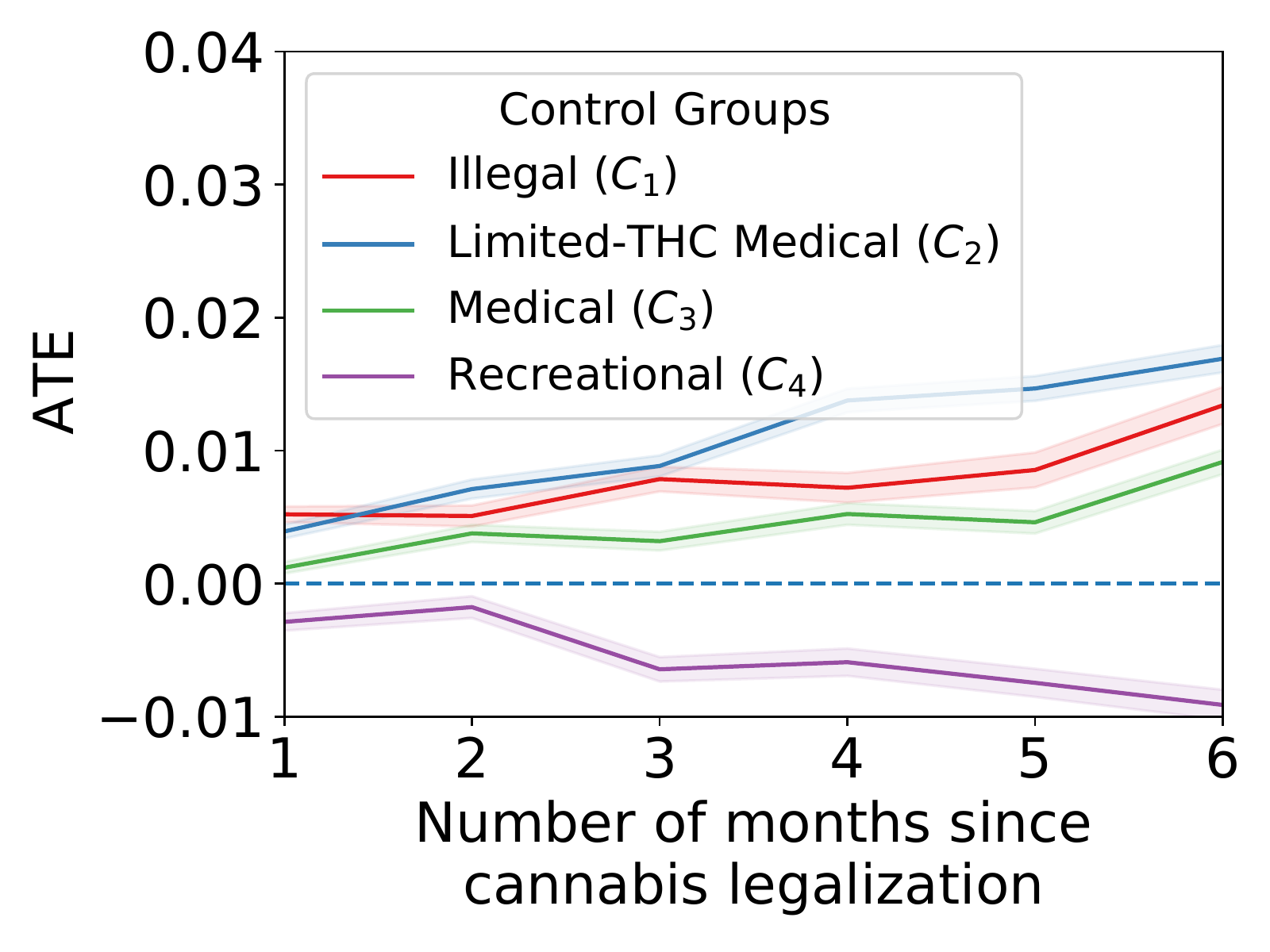}
        \caption{IPTW with LR (no retweets)}
        \label{fig:noretweet_cal}
     \end{subfigure}

        \caption{(a-e): Sensitivity analysis of Average Treatment Effect (ATE) for matching methods when treatment state is California and retweets are included. (f): ATE estimation with IPTW-LR (best model) when retweets are excluded. The x-axis shows the number of months since recreational cannabis legalization in California.}
        \label{fig:sen_effect}
\end{figure}

\textbf{Sensitivity analysis.} Next, we show the sensitivity of effect estimates to the choice of matching model: IPTW with LR (chosen), IPTW with GBM, PSM with LR, PSM with GBM, and NNM with cosine similarity. 
Figure \ref{fig:sen_balance} depicts the covariate balance obtained after running five matching models for California as the treatment state. 
All the matching models improve the covariate balance after matching with IPTW-LR having the lowest ASMD.
Figure \ref{fig:sen_effect} shows the estimated causal effects for the treatment state of California and the same five models described above. We observe a nearly consistent magnitude of ATE for the states with illegal cannabis across all the matching methods. There is some variation in the estimation of ATE for control groups other than $C_1$. IPTW method has tighter confidence intervals compared to other matching models. Also, the choice of the propensity score model is less sensitive in IPTW compared to PSM. 

Figures \ref{fig:tw_iptw_lr_bal} and \ref{fig:noretweet_cal} show the IPTW-LR covariate balance and causal effect estimated for California when all retweets are discarded. The covariate balance before and after matching is consistent with the settings when retweets are included. Although the effects are smaller compared to the estimate including retweets for control groups $C_1$ and $C_4$, the generic trends of those control groups across the three treatment states are still preserved. The trends for $C_2$ and $C_3$ are fairly consistent with the estimate including retweets. The population size for Massachusetts and Vermont is too small for the analysis when retweets are excluded.

\section{Potential Limitations}
One of the limitations of our work is associated with the noise and bias in Twitter data. Although studies have suggested correspondence between Twitter activities and real use of e-cigarettes and cannabis among young adults~\cite{cabrera-nguyen-jsad16,unger-pm18}, Twitter activities are noisy proxies to the real-world phenomenon. Moreover, selection bias can potentially be introduced in our pipeline of data collection, location filtering, and personal tweet selection. %Figure \ref{fig:state_dist} shows the distribution of users considered in our study roughly follows the population size rank of the United States. Also,  
Similarly, the errors in the stance detection models can act as a potential measurement bias for the downstream task of causal effect estimation which is why we use probabilistic classifiers to capture measurement errors as the variance of the causal effect estimate.

Another limitation is related to the causal inference assumptions. For example, there may be latent confounders that violate the ignorability assumption of propensity score matching. However, if we assume that state assignment is close to randomized, i.e., users could be living in any of the states, then this assumption makes sense. Then the balancing approach addresses inherent differences between specific users and thus reduces the confounding bias assuming the unobserved confounders are often associated with the observed covariates~\cite{stuart-sc10}. Another possible concern is the violation of the SUTVA assumption, where the treatment can ``spill over" to the control individuals. An example in our context can be a user from a control state tweeting in support of recreational cannabis legalization in the treatment state. Even in presence of spillover, the estimated ATE, the difference of pro-cannabis tweet initiation rate, is likely an underestimation of true ATE. %Similarly, there may be cases of contagion where the outcome of a unit can affect other units' outcome. 
Some potential future directions include studying the effect of network interference and the generalizability of the results to other states.

% Although our findings are based on the trends at three different points in time across three states with disparate demography, more study is needed to strengthen the generalizability of the results.

% \input{7-conclusion}
\section{Conclusion}
In this work, we show that cannabis legalization likely has an impact on cannabis uptake for vaping users. Using a novel partially-annotated Twitter dataset, we leverage weakly supervised learning and transfer learning for personal tweet selection as well as e-cigarette and cannabis stance detection. We perform sensitivity analysis of different causal model choices on the estimation. We discover users with a positive attitude toward e-cigarettes in states with legalized recreational cannabis are much more likely to develop a pro-cannabis attitude and potentially start cannabis vaping than users in states where cannabis is illegal.

\section*{Acknowledgements}

We thank the anonymous reviewers for their valuable feedback. We also thank Cornelia Caragea and Barbara Di Eugenio for their help with the annotation design and evaluation. This material is based on research sponsored in part by the National Institute on Drug Abuse (NIDA) under grant number 1 R01 DA051157 and the Defense Advanced Research Projects Agency (DARPA) under contract number \\HR00111990114. The views and conclusions contained herein are those of the authors and should not be interpreted as necessarily representing the official policies, either expressed or implied, of DARPA or the U.S. Government. The U.S. Government is authorized to reproduce and
distribute reprints for governmental purposes notwithstanding any copyright annotation therein.

\bibliography{references}

\begin{thebibliography}{58}
\providecommand{\natexlab}[1]{#1}
\providecommand{\url}[1]{\texttt{#1}}
\providecommand{\urlprefix}{URL }
\expandafter\ifx\csname urlstyle\endcsname\relax
  \providecommand{\doi}[1]{doi:\discretionary{}{}{}#1}\else
  \providecommand{\doi}{doi:\discretionary{}{}{}\begingroup
  \urlstyle{rm}\Url}\fi

\bibitem[{Abebe et~al.(2020)Abebe, Giorgi, Tedijanto, Buffone, and
  Schwartz}]{abebe-www20}
Abebe, R.; Giorgi, S.; Tedijanto, A.; Buffone, A.; and Schwartz, H. A.~A. 2020.
\newblock Quantifying Community Characteristics of Maternal Mortality Using
  Social Media.
\newblock In \emph{WWW}, 2976--2983.

\bibitem[{Agrawal, Budney, and Lynskey(2012)}]{agrawal-a12}
Agrawal, A.; Budney, A.~J.; and Lynskey, M.~T. 2012.
\newblock The co-occurring use and misuse of cannabis and tobacco: a review.
\newblock \emph{Addiction} 107(7): 1221--1233.

\bibitem[{Aldayel and Magdy(2019)}]{aldayel-icsi19}
Aldayel, A.; and Magdy, W. 2019.
\newblock Assessing Sentiment of the Expressed Stance on Social Media.
\newblock In \emph{International Conference on Social Informatics}, 277--286.
  Springer.

\bibitem[{AlDayel and Magdy(2020)}]{aldayel-arxiv20}
AlDayel, A.; and Magdy, W. 2020.
\newblock Stance Detection on Social Media: State of the Art and Trends.
\newblock \emph{arXiv preprint arXiv:2006.03644} .

\bibitem[{Altenburger et~al.(2017)Altenburger, De, Frazier, Avteniev, and
  Hamilton}]{altenburger-icwsm17}
Altenburger, K.~M.; De, R.; Frazier, K.; Avteniev, N.; and Hamilton, J. 2017.
\newblock Are there gender differences in professional self-promotion? an
  empirical case study of linkedin profiles among recent mba graduates.
\newblock In \emph{ICWSM}.

\bibitem[{Audrain-McGovern et~al.(2018)Audrain-McGovern, Stone,
  Barrington-Trimis, Unger, and Leventhal}]{audrain-pediatrics18}
Audrain-McGovern, J.; Stone, M.~D.; Barrington-Trimis, J.; Unger, J.~B.; and
  Leventhal, A.~M. 2018.
\newblock Adolescent e-cigarette, hookah, and conventional cigarette use and
  subsequent marijuana use.
\newblock \emph{Pediatrics} 142(3): e20173616.

\bibitem[{Austin(2016)}]{austin-sm16}
Austin, P.~C. 2016.
\newblock Variance estimation when using inverse probability of treatment
  weighting (IPTW) with survival analysis.
\newblock \emph{Statistics in medicine} 35(30): 5642--5655.

\bibitem[{Barrington-Trimis et~al.(2015)Barrington-Trimis, Berhane, Unger,
  Cruz, Huh, Leventhal, Urman, Wang, Howland, Gilreath
  et~al.}]{barrington-pediatrics15}
Barrington-Trimis, J.~L.; Berhane, K.; Unger, J.~B.; Cruz, T.~B.; Huh, J.;
  Leventhal, A.~M.; Urman, R.; Wang, K.; Howland, S.; Gilreath, T.~D.; et~al.
  2015.
\newblock Psychosocial factors associated with adolescent electronic cigarette
  and cigarette use.
\newblock \emph{Pediatrics} 136(2): 308--317.

\bibitem[{Barrington-Trimis et~al.(2016)Barrington-Trimis, Berhane, Unger,
  Cruz, Urman, Chou, Howland, Wang, Pentz, Gilreath et~al.}]{barrington-jah16}
Barrington-Trimis, J.~L.; Berhane, K.; Unger, J.~B.; Cruz, T.~B.; Urman, R.;
  Chou, C.~P.; Howland, S.; Wang, K.; Pentz, M.~A.; Gilreath, T.~D.; et~al.
  2016.
\newblock The e-cigarette social environment, e-cigarette use, and
  susceptibility to cigarette smoking.
\newblock \emph{Journal of Adolescent Health} 59(1): 75--80.

\bibitem[{Borodovsky et~al.(2016)Borodovsky, Crosier, Lee, Sargent, and
  Budney}]{borodovsky-ijdp16}
Borodovsky, J.~T.; Crosier, B.~S.; Lee, D.~C.; Sargent, J.~D.; and Budney,
  A.~J. 2016.
\newblock Smoking, vaping, eating: Is legalization impacting the way people use
  cannabis?
\newblock \emph{International Journal of Drug Policy} 36: 141--147.

\bibitem[{Borodovsky et~al.(2017)Borodovsky, Lee, Crosier, Gabrielli, Sargent,
  and Budney}]{borodovsky-dad17}
Borodovsky, J.~T.; Lee, D.~C.; Crosier, B.~S.; Gabrielli, J.~L.; Sargent,
  J.~D.; and Budney, A.~J. 2017.
\newblock US cannabis legalization and use of vaping and edible products among
  youth.
\newblock \emph{Drug and alcohol dependence} 177: 299--306.

\bibitem[{Broniatowski et~al.(2018)Broniatowski, Jamison, Qi, AlKulaib, Chen,
  Benton, Quinn, and Dredze}]{broniatowski-ajph18}
Broniatowski, D.~A.; Jamison, A.~M.; Qi, S.; AlKulaib, L.; Chen, T.; Benton,
  A.; Quinn, S.~C.; and Dredze, M. 2018.
\newblock Weaponized health communication: Twitter bots and Russian trolls
  amplify the vaccine debate.
\newblock \emph{American journal of public health} 108(10): 1378--1384.

\bibitem[{Cabrera-Nguyen et~al.(2016)Cabrera-Nguyen, Cavazos-Rehg, Krauss,
  Bierut, and Moreno}]{cabrera-nguyen-jsad16}
Cabrera-Nguyen, E.~P.; Cavazos-Rehg, P.; Krauss, M.; Bierut, L.~J.; and Moreno,
  M.~A. 2016.
\newblock Young adults’ exposure to alcohol-and marijuana-related content on
  Twitter.
\newblock \emph{Journal of studies on alcohol and drugs} 77(2): 349--353.

\bibitem[{Cavazos-Rehg et~al.(2014)Cavazos-Rehg, Krauss, Grucza, and
  Bierut}]{cavazos-jmir14}
Cavazos-Rehg, P.; Krauss, M.; Grucza, R.; and Bierut, L. 2014.
\newblock Characterizing the followers and tweets of a marijuana-focused
  Twitter handle.
\newblock \emph{JMIR} 16(6): e157.

\bibitem[{Chatham-Stephens et~al.(2019)Chatham-Stephens, Roguski, Jang, Cho,
  Jatlaoui, Kabbani, Glidden, Ussery, Trivers, Evans et~al.}]{chatham-mmwr19}
Chatham-Stephens, K.; Roguski, K.; Jang, Y.; Cho, P.; Jatlaoui, T.~C.; Kabbani,
  S.; Glidden, E.; Ussery, E.~N.; Trivers, K.~F.; Evans, M.~E.; et~al. 2019.
\newblock Characteristics of hospitalized and nonhospitalized patients in a
  nationwide outbreak of e-cigarette, or vaping, product use--associated lung
  injury—United States, November 2019.
\newblock \emph{Morbidity and Mortality Weekly Report} 68(46): 1076.

\bibitem[{Conway(2020)}]{conway-statista20}
Conway, J. 2020.
\newblock Vaping market share in the United States in 2020, by brand.
\newblock
  \url{https://www.statista.com/statistics/1096995/vaping-market-share-us-by-brand}.
\newblock Accessed: 2020-09-08.

\bibitem[{Coppersmith, Dredze, and Harman(2014)}]{coppersmith-acl14}
Coppersmith, G.; Dredze, M.; and Harman, C. 2014.
\newblock Quantifying mental health signals in Twitter.
\newblock In \emph{Workshop on computational linguistics and clinical
  psychology: From linguistic signal to clinical reality}, 51--60.

\bibitem[{Cullen et~al.(2018)Cullen, Ambrose, Gentzke, Apelberg, Jamal, and
  King}]{cullen2018notes}
Cullen, K.~A.; Ambrose, B.~K.; Gentzke, A.~S.; Apelberg, B.~J.; Jamal, A.; and
  King, B.~A. 2018.
\newblock Notes from the field: use of electronic cigarettes and any tobacco
  product among middle and high school students—United States, 2011--2018.
\newblock \emph{Morbidity and Mortality Weekly Report} 67(45): 1276.

\bibitem[{Dai et~al.(2018)Dai, Catley, Richter, Goggin, and
  Ellerbeck}]{dai-pediatrics18}
Dai, H.; Catley, D.; Richter, K.~P.; Goggin, K.; and Ellerbeck, E.~F. 2018.
\newblock Electronic cigarettes and future marijuana use: a longitudinal study.
\newblock \emph{Pediatrics} 141(5): e20173787.

\bibitem[{De~Choudhury and K{\i}c{\i}man(2017)}]{de-icwsm17}
De~Choudhury, M.; and K{\i}c{\i}man, E. 2017.
\newblock The language of social support in social media and its effect on
  suicidal ideation risk.
\newblock In \emph{ICWSM}, volume 2017, 32.

\bibitem[{Dos~Reis and Culotta(2015)}]{dos-aaai15}
Dos~Reis, V.~L.; and Culotta, A. 2015.
\newblock Using matched samples to estimate the effects of exercise on mental
  health from twitter.
\newblock In \emph{AAAI}, 182--188.

\bibitem[{Giroud et~al.(2015)Giroud, De~Cesare, Berthet, Varlet, Concha-Lozano,
  and Favrat}]{giroud-ijerph15}
Giroud, C.; De~Cesare, M.; Berthet, A.; Varlet, V.; Concha-Lozano, N.; and
  Favrat, B. 2015.
\newblock E-cigarettes: a review of new trends in cannabis use.
\newblock \emph{International Journal of Environmental Research and Public
  Health} 12(8): 9988--10008.

\bibitem[{Hajek et~al.(2019)Hajek, Phillips-Waller, Przulj, Pesola,
  Myers~Smith, Bisal, Li, Parrott, Sasieni, Dawkins et~al.}]{hajek-nejm19}
Hajek, P.; Phillips-Waller, A.; Przulj, D.; Pesola, F.; Myers~Smith, K.; Bisal,
  N.; Li, J.; Parrott, S.; Sasieni, P.; Dawkins, L.; et~al. 2019.
\newblock A randomized trial of e-cigarettes versus nicotine-replacement
  therapy.
\newblock \emph{New England Journal of Medicine} 380(7): 629--637.

\bibitem[{Harris et~al.(2014)Harris, Moreland-Russell, Choucair, Mansour,
  Staub, and Simmons}]{harris-jmir14}
Harris, J.~K.; Moreland-Russell, S.; Choucair, B.; Mansour, R.; Staub, M.; and
  Simmons, K. 2014.
\newblock Tweeting for and against public health policy: response to the
  Chicago Department of Public Health's electronic cigarette Twitter campaign.
\newblock \emph{JMIR} 16(10): e238.

\bibitem[{Hatchard et~al.(2019)Hatchard, Quariguasi Frota~Neto, Vasilakis, and
  Evans-Reeves}]{hatchard-plos19}
Hatchard, J.~L.; Quariguasi Frota~Neto, J.; Vasilakis, C.; and Evans-Reeves,
  K.~A. 2019.
\newblock Tweeting about public health policy: Social media response to the UK
  Government’s announcement of a Parliamentary vote on draft standardised
  packaging regulations.
\newblock \emph{PloS one} 14(2): e0211758.

\bibitem[{Hutto and Gilbert(2014)}]{hutto-aaai14}
Hutto, C.~J.; and Gilbert, E. 2014.
\newblock Vader: A parsimonious rule-based model for sentiment analysis of
  social media text.
\newblock In \emph{ICWSM}.

\bibitem[{Jackler et~al.(2019)Jackler, Chau, Getachew, Whitcomb,
  Lee-Heidenreich, Bhatt, Kim-O’Sullivan, Hoffman, Jackler, and
  Ramamurthi}]{jackler-srita19}
Jackler, R.~K.; Chau, C.; Getachew, B.~D.; Whitcomb, M.~M.; Lee-Heidenreich,
  J.; Bhatt, A.~M.; Kim-O’Sullivan, S.~H.; Hoffman, Z.~A.; Jackler, L.~M.;
  and Ramamurthi, D. 2019.
\newblock JUUL advertising over its first three years on the market.
\newblock \emph{Stanford Research into the Impact of Tobacco Advertising White
  Paper} .

\bibitem[{Jiang, Calix, and Gupta(2016)}]{jiang-aclworkshop16}
Jiang, K.; Calix, R.; and Gupta, M. 2016.
\newblock Construction of a personal experience tweet corpus for health
  surveillance.
\newblock In \emph{Proceedings of the 15th workshop on biomedical NLP},
  128--135.

\bibitem[{Jiang et~al.(2018)Jiang, Feng, Song, Calix, Gupta, and
  Bernard}]{jiang-bmc18}
Jiang, K.; Feng, S.; Song, Q.; Calix, R.~A.; Gupta, M.; and Bernard, G.~R.
  2018.
\newblock Identifying tweets of personal health experience through word
  embedding and LSTM neural network.
\newblock \emph{BMC bioinformatics} 19(8): 210.

\bibitem[{Johnston et~al.(2020)Johnston, Miech, O’Malley, J.~G.~Bachman, and
  Patrick}]{mtf-2020}
Johnston, L.~D.; Miech, R.~A.; O’Malley, P.~M.; J.~G.~Bachman, J. E.~S.; and
  Patrick, M.~E. 2020.
\newblock Monitoring the Future national survey results on drug use 1975-2019:
  Overview, key findings on adolescent drug use.
\newblock \emph{Institute for Social Research, University of Michigan} .

\bibitem[{K{\i}c{\i}man, Counts, and Gasser(2018)}]{kiciman-icwsm18}
K{\i}c{\i}man, E.; Counts, S.; and Gasser, M. 2018.
\newblock Using longitudinal social media analysis to understand the effects of
  early college alcohol use.
\newblock \emph{ICWSM} .

\bibitem[{Kim et~al.(2017)Kim, Miano, Chew, Eggers, and
  Nonnemaker}]{kim-jmir17}
Kim, A.; Miano, T.; Chew, R.; Eggers, M.; and Nonnemaker, J. 2017.
\newblock Classification of Twitter users who tweet about e-cigarettes.
\newblock \emph{JMIR public health and surveillance} 3(3): e63.

\bibitem[{King and Nielsen(2019)}]{king-pa19}
King, G.; and Nielsen, R.~A. 2019.
\newblock Why propensity scores should not be used for matching.
\newblock \emph{Political Analysis} 27(4).

\bibitem[{Kong et~al.(2014)Kong, Schneider, Swayamdipta, Bhatia, Dyer, and
  Smith}]{kong-emnlp14}
Kong, L.; Schneider, N.; Swayamdipta, S.; Bhatia, A.; Dyer, C.; and Smith,
  N.~A. 2014.
\newblock A dependency parser for tweets.
\newblock In \emph{EMNLP}, 1001--1012.

\bibitem[{Lemyre, Poliakova, and B{\'e}langer(2019)}]{lemyre-sum19}
Lemyre, A.; Poliakova, N.; and B{\'e}langer, R.~E. 2019.
\newblock The relationship between tobacco and cannabis use: a review.
\newblock \emph{Substance Use \& Misuse} 54(1): 130--145.

\bibitem[{Liu(2012)}]{liu-slhlt12}
Liu, B. 2012.
\newblock Sentiment analysis and opinion mining.
\newblock \emph{Synthesis lectures on human language technologies} 5(1):
  1--167.

\bibitem[{Liu et~al.(2019)Liu, Shi, Wu, Thomas, Symul, Pierson, and
  Leskovec}]{liu-www19}
Liu, B.; Shi, S.; Wu, Y.; Thomas, D.; Symul, L.; Pierson, E.; and Leskovec, J.
  2019.
\newblock Predicting pregnancy using large-scale data from a women's health
  tracking mobile application.
\newblock In \emph{WWW}, 2999--3005.

\bibitem[{Metaxas et~al.(2015)Metaxas, Mustafaraj, Wong, Zeng, O'Keefe, and
  Finn}]{metaxas-icwsm15}
Metaxas, P.; Mustafaraj, E.; Wong, K.; Zeng, L.; O'Keefe, M.; and Finn, S.
  2015.
\newblock What do retweets indicate? Results from user survey and meta-review
  of research.
\newblock In \emph{ICWSM}, volume~9.

\bibitem[{Morean et~al.(2017)Morean, Lipshie, Josephson, and
  Foster}]{morean-sum17}
Morean, M.~E.; Lipshie, N.; Josephson, M.; and Foster, D. 2017.
\newblock Predictors of adult e-cigarette users vaporizing cannabis using
  e-cigarettes and vape-pens.
\newblock \emph{Substance Use \& Misuse} 52(8): 974--981.

\bibitem[{Pan and Yang(2009)}]{pan-kde09}
Pan, S.~J.; and Yang, Q. 2009.
\newblock A survey on transfer learning.
\newblock \emph{IEEE Transactions on knowledge and data engineering} 22(10):
  1345--1359.

\bibitem[{Park et~al.(2020)Park, Kwak, Song, and Cha}]{park-icwsm20}
Park, K.; Kwak, H.; Song, H.; and Cha, M. 2020.
\newblock “Trust Me, I Have a Ph. D.”: A Propensity Score Analysis on the
  Halo Effect of Disclosing One's Offline Social Status in Online Communities.
\newblock In \emph{ICWSM}, volume~14, 534--544.

\bibitem[{Pennington, Socher, and Manning(2014)}]{pennington-emnlp14}
Pennington, J.; Socher, R.; and Manning, C.~D. 2014.
\newblock Glove: Global vectors for word representation.
\newblock In \emph{EMNLP}, 1532--1543.

\bibitem[{Polosa et~al.(2019)Polosa, O’Leary, Tashkin, Emma, and
  Caruso}]{polosa-errm19}
Polosa, R.; O’Leary, R.; Tashkin, D.; Emma, R.; and Caruso, M. 2019.
\newblock The effect of e-cigarette aerosol emissions on respiratory health: a
  narrative review.
\newblock \emph{Expert review of respiratory medicine} 13(9): 899--915.

\bibitem[{Rajagopal et~al.(2013)Rajagopal, Cambria, Olsher, and
  Kwok}]{rajagopal-www13}
Rajagopal, D.; Cambria, E.; Olsher, D.; and Kwok, K. 2013.
\newblock A graph-based approach to commonsense concept extraction and semantic
  similarity detection.
\newblock In \emph{WWW}, 565--570.

\bibitem[{Ratner et~al.(2017)Ratner, Bach, Ehrenberg, Fries, Wu, and
  R{\'e}}]{ratner-vldb17}
Ratner, A.; Bach, S.~H.; Ehrenberg, H.; Fries, J.; Wu, S.; and R{\'e}, C. 2017.
\newblock Snorkel: Rapid training data creation with weak supervision.
\newblock In \emph{VLDB}, volume~11, 269.

\bibitem[{Rubin(1974)}]{rubin-jep74}
Rubin, D.~B. 1974.
\newblock Estimating causal effects of treatments in randomized and
  nonrandomized studies.
\newblock \emph{Journal of Educational Psychology} 66(5): 688.

\bibitem[{Shadish et~al.(2002)Shadish, Cook, Campbell et~al.}]{shadish-book02}
Shadish, W.~R.; Cook, T.~D.; Campbell, D.~T.; et~al. 2002.
\newblock \emph{Experimental and quasi-experimental designs for generalized
  causal inference}.
\newblock Boston: Houghton Mifflin.

\bibitem[{Shahid and Zheleva(2019)}]{shahid-why19}
Shahid, U.; and Zheleva, E. 2019.
\newblock Empirical Study of Model Dependence in Counterfactual Learning from
  Networks.
\newblock In \emph{AAAI Spring Symposium on Beyond Curve Fitting: Causation,
  Counterfactuals, and Imagination-based AI (AAAI-WHY)}.

\bibitem[{Smart and Pacula(2019)}]{smart-ajdaa19}
Smart, R.; and Pacula, R.~L. 2019.
\newblock Early evidence of the impact of cannabis legalization on cannabis
  use, cannabis use disorder, and the use of other substances: findings from
  state policy evaluations.
\newblock \emph{The American journal of drug and alcohol abuse} 45(6):
  644--663.

\bibitem[{Stuart(2010)}]{stuart-sc10}
Stuart, E.~A. 2010.
\newblock Matching methods for causal inference: A review and a look forward.
\newblock \emph{Statistical science: a review journal of the Institute of
  Mathematical Statistics} 25(1): 1.

\bibitem[{Tian and Chunara(2020)}]{tian-aaai20}
Tian, Y.; and Chunara, R. 2020.
\newblock Quasi-Experimental Designs for Assessing Response on Social Media to
  Policy Changes.
\newblock In \emph{ICWSM}, volume~14, 671--682.

\bibitem[{Unger et~al.(2018)Unger, Urman, Cruz, Majmundar, Barrington-Trimis,
  Pentz, and McConnell}]{unger-pm18}
Unger, J.~B.; Urman, R.; Cruz, T.~B.; Majmundar, A.; Barrington-Trimis, J.;
  Pentz, M.~A.; and McConnell, R. 2018.
\newblock Talking about tobacco on Twitter is associated with tobacco product
  use.
\newblock \emph{Preventive medicine} 114: 54--56.

\bibitem[{{U.S. Department of Health and Human
  Services}(2016)}]{usdhhs-nccdphp16}
{U.S. Department of Health and Human Services}. 2016.
\newblock E-Cigarette Use Among Youth and Young Adults. A Report of the Surgeon
  General.
\newblock \emph{U.S. Department of HHS, CDC, National Center for Chronic
  Disease Prevention and Health Promotion, Office on Smoking and Health}
  1--275.

\bibitem[{{U.S. Food \& Drug Administration}(2018)}]{fda-online18}
{U.S. Food \& Drug Administration}. 2018.
\newblock Tobacco 21.
\newblock
  \url{https://www.fda.gov/tobacco-products/retail-sales-tobacco-products/tobacco-21}.
\newblock Accessed: 2020-07-18.

\bibitem[{Walker et~al.(2020)Walker, Parag, Verbiest, Laking, Laugesen, and
  Bullen}]{walker-lrm20}
Walker, N.; Parag, V.; Verbiest, M.; Laking, G.; Laugesen, M.; and Bullen, C.
  2020.
\newblock Nicotine patches used in combination with e-cigarettes (with and
  without nicotine) for smoking cessation: a pragmatic, randomised trial.
\newblock \emph{The Lancet Respiratory Medicine} 8(1): 54--64.

\bibitem[{Wilson, Wiebe, and Hoffmann(2005)}]{wilson-emnlp05}
Wilson, T.; Wiebe, J.; and Hoffmann, P. 2005.
\newblock Recognizing contextual polarity in phrase-level sentiment analysis.
\newblock In \emph{HLT/EMNLP}, 347--354.

\bibitem[{Wojcik and Hughes(2019)}]{TwitterAge}
Wojcik, S.; and Hughes, A. 2019.
\newblock Sizing Up Twitter Users.
\newblock
  \url{https://www.pewresearch.org/internet/2019/04/24/sizing-up-twitter-users/}.
\newblock Accessed: 2020-07-18.

\bibitem[{Young, Padwa, and Bonar(2019)}]{young-fph19}
Young, S.; Padwa, H.; and Bonar, E. 2019.
\newblock Social Big Data as a Tool for Understanding and Predicting the Impact
  of Cannabis Legalization.
\newblock \emph{Frontiers in public health} 7: 274.

\end{thebibliography}

\end{document}